\documentclass[a4paper]{article}
\usepackage{etoolbox}

\usepackage{amsmath,amsthm}
\usepackage{mathtools}
\usepackage[dvipsnames,table]{xcolor}

\usepackage[numbers]{natbib}
\usepackage{authblk}

\RequirePackage{crop}
\RequirePackage{graphicx}
\usepackage[labelfont=it]{caption}
\RequirePackage{array}
\RequirePackage{color}
\RequirePackage{amssymb}
\RequirePackage{flushend}
\RequirePackage{stfloats}
\RequirePackage[figuresright]{rotating}
\RequirePackage{chngpage}
\RequirePackage{totcount}
\RequirePackage{fix-cm}
\RequirePackage{hyperref}
\usepackage{doi}
\usepackage{tabularx, booktabs, multirow}
\usepackage{makecell}
\usepackage{placeins}

\definecolor{linkcolor}{HTML}{991408}  
\definecolor{citecolor}{HTML}{2E7E2A}  
\definecolor{filecolor}{HTML}{131877}  
\definecolor{menucolor}{HTML}{727500}  
\definecolor{runcolor} {HTML}{137776}  
\definecolor{urlcolor} {HTML}{0a2bbf}  
\hypersetup{colorlinks=true,linkcolor=linkcolor,citecolor=citecolor,filecolor=filecolor,menucolor=menucolor,runcolor=runcolor,urlcolor=urlcolor}

\newtoggle{arxiv}
\toggletrue{arxiv}
\iftoggle{arxiv}{
  \usepackage[parfill]{parskip}

  \usepackage[%
    a4paper,
    inner=25mm,
    outer=25mm,
    top=25mm,
    bottom=25mm,
    marginparsep=5mm,
    marginparwidth=40mm,
  ]{geometry}

  \newlength{\defbaselineskip}
  \setlength{\defbaselineskip}{\baselineskip}
  \setlength{\parskip}{6pt}%
  \setlength{\parindent}{0pt}%

  \RequirePackage[T1]{fontenc}
  \RequirePackage[tt=false, type1=true]{libertine}
  \RequirePackage[varqu]{zi4}
  \RequirePackage[libertine]{newtxmath}
}{}

\usepackage{tikz}
\usepackage{float}
\usepackage{subcaption}

\usepackage{pifont}

\usepackage{comment}

\newcommand{\na}[1]{{\color{darkgray}{N/A}}}

\newcommand{\appref}[1]{\hyperref[#1]{Appendix~\ref*{#1}}}

\usepackage{fancyhdr}
\usepackage{multicol}
\makeatletter
\renewcommand\@author{
    \AB@authlist\\[\affilsep]
    \begin{minipage}{0.9\textwidth}
    \begin{multicols}{2}
        \raggedright
        \AB@affillist
    \end{multicols}
    \end{minipage}
    }
\makeatother

\title{BarcodeMAE+: Rethinking Masked Pretraining and Global Representations for DNA Barcode Foundation Models}

\author[1,$\ast$,$\dagger$]{Monireh Safari}
\author[1,$\dagger$]{Pablo Millan Arias}
\author[4]{Scott C. Lowe}
\author[1]{Lila Kari}
\author[3,5]{Angel X. Chang}
\author[2,4]{Graham W. Taylor}

\affil[1]{Cheriton School of Computer Science, University of Waterloo, Ontario, Canada}
\affil[2]{Department of Interdisciplinary Engineering, University of Guelph, Ontario, Canada}
\affil[3]{School of Computing Science, Simon Fraser University, British Columbia, Canada}
\affil[4]{Vector Institute, Ontario, Canada}
\affil[5]{Alberta Machine Intelligence Institute (Amii), Alberta, Canada}
\affil[$\ast$]{Corresponding author: \href{mailto:monireh.safari@uwaterloo.ca}{monireh.safari@uwaterloo.ca}}
\affil[$\dagger$]{These authors contributed equally to this work.}

\begin{document}

\maketitle
\begin{abstract}
\textbf{Background:}
Many DNA foundation models are pretrained by masking parts of a sequence and asking the model to reconstruct them. Standard masked pretraining exposes the encoder to special \texttt{[MASK]} tokens that are absent at inference, creating a mismatch between training and downstream use. The role of an explicit global sequence representation such as a \texttt{[CLS]} token and how it should be trained also remain poorly understood for DNA barcodes.

\textbf{Results:}
We introduce BarcodeMAE+ and study model architecture, global \texttt{[CLS]} representation, and auxiliary pretraining objectives across arthropod COI (BIOSCAN-5M) and fungal ITS (UNITE+INSD) barcodes. Across both barcode regions, the encoder-decoder MAE-LM architecture outperforms its matched encoder-only counterpart in nearly all evaluated configurations, supporting MAE-LM as an effective architectural design for DNA barcode foundation models. A trained global \texttt{[CLS]} representation provides substantial additional gains: on BIOSCAN-5M, \texttt{[CLS]} accuracy increases from 47.53\% without an auxiliary objective to 80.65\% with cross-entropy genus classification. The best auxiliary objective is region-dependent: cross-entropy performs best on BIOSCAN-5M, whereas pairwise same-genus classification performs best on UNITE+INSD, reaching 73.19\% on Yeast and 63.07\% on Filamentous Fungi. BarcodeMAE+ outperforms published DNA foundation model baselines on BIOSCAN-5M and achieves the highest Yeast accuracy among the evaluated UNITE+INSD baselines using frozen encoder representations. Similarity-weighted softmax KNN voting further stabilizes accuracy as neighbourhood size increases.

\textbf{Conclusions:}
Encoder-decoder masked pretraining and an explicitly trained global representation are strong design choices for DNA barcode foundation models, while the optimal objective for learning that representation depends on the biological domain.
\end{abstract}


\section{Introduction}
\label{sec:intro}
\vspace{-4pt}
DNA barcodes are short, standardized genomic regions that vary little within a species and differ clearly between species. This makes DNA barcoding a successful species-level specimen identification tool that assigns specimens to species from DNA barcodes, offering a faster alternative to morphological identification~\citep{hebert2003biological}. The standard barcode region depends on the taxonomic group: for animals it is a
658-bp fragment of the Cytochrome c Oxidase Subunit I (COI) gene,  while for fungi it is the Internal Transcribed Spacer (ITS) region. Beyond the identification of existing species, barcodes support the discovery of new species and help group specimens into taxonomic units even before formal species boundaries are defined, making them central to large-scale biodiversity monitoring.

DNA foundation models are now widely used to analyze genomic sequences, spanning
architectures such as BERT-like transformers~\citep{devlin2019bert,ji2021dnabert,millan2023barcodebert,zhou2024dnabertS,dalla2024nucleotide,zhou2023dnabert2,fishman2025genalm},
state-space models~\citep{poli2023hyena,nguyen2024hyenadna,schiff2024caduceus,gao2024barcodemamba,gao2025barcodemambaplus},
and convolutional neural networks~\citep{benegas2023dna}. These architectures are typically paired with one of two broad pretraining objectives: bidirectional masked language modelling, which predicts each hidden token from its surrounding context, or causal, autoregressive next-token prediction, which predicts each token from only the previous ones in the sequence. Both paradigms have proven effective, and the choice between them is often driven by the downstream requirements: autoregressive pretraining is a natural choice for sequence generation and extrapolation of long-range dependencies, whereas bidirectional pretraining produces stronger encoders for understanding tasks, such as classification. By letting every position attend to its full context, each output token embedding can encode information from the entire sequence, resulting in more powerful aggregated representations for downstream tasks that rely on a frozen, sequence-level embedding, such as retrieval and clustering. Masked language modelling (MLM)~\citep{devlin2019bert} remains one of the most widely adopted pretraining strategies for DNA foundation models used in this way, and it is the one we build on in this paper.

During MLM pretraining, a random subset of input tokens is replaced with a special \texttt{[MASK]} placeholder, and the model is trained to predict the original tokens at those positions from the surrounding sequence. This objective has been shown to learn effective sequence representations for downstream tasks such as specimen identification and novel-species discovery. However, MLM introduces a well-known distributional shift between pretraining and inference, as \texttt{[MASK]} tokens are in the input sequence during pretraining but not at inference~\citep{meng2024maelm}. As a result, part of the encoder's capacity is devoted to processing \texttt{[MASK]} embeddings that are not used downstream, potentially reducing the quality of the learned representations. More broadly, recent work on text encoders challenges the assumption that improved token reconstruction necessarily produces better downstream representations. As standard BERT-style models scale, reconstruction perplexity can improve while frozen-probe performance deteriorates, a misalignment attributed to the architecture directly coupling encoder representations to the token-reconstruction loss~\citep{dervishi2026crossbert}.

This is particularly important in biodiversity informatics, where foundation models are often used as frozen feature extractors for tasks such as KNN-based specimen identification and zero-shot species clustering, without task-specific fine-tuning. BarcodeMAE~\citep{safari2025barcodeMAE}, which we previously introduced in a preliminary workshop paper, explored this approach for DNA barcode modelling by adapting the MAE-LM framework~\citep{meng2024maelm} to arthropod COI sequences. By removing masked tokens from the encoder during pretraining, BarcodeMAE achieved substantial improvements over standard MLM on BIOSCAN-5M~\citep{gharaee2024bioscan5m}, a large-scale reference library of arthropod DNA barcodes. In this paper, we extend the initial BarcodeMAE workshop paper by evaluating a second barcode region, the fungal ITS marker (UNITE+INSD), and by systematically investigating design choices for aggregating token-level representations into a global sequence embedding.

A strong global sequence representation is an important component of DNA foundation models, as many downstream applications require a compact embedding that captures information from the entire sequence. 
There are different strategies to obtain this aggregate embedding.  
A simple approach is to take the average (mean) of the embeddings of the sequence tokens.
Another common approach is to prepend a learnable classification token, \texttt{[CLS]}, that attends to the sequence and serves as its global representation. However, simply introducing a \texttt{[CLS]} token to the input does not guarantee that it will learn to output an informative sequence-level embedding. Under standard MLM, the reconstruction objective is applied only to masked token predictions and does not provide an explicit objective for the \texttt{[CLS]} representation. 
This raises two closely related design questions in the context of foundation models for DNA barcoding: whether an explicit global token improves downstream performance, and which training objective should be used to shape that representation. Auxiliary objectives, such as pairwise same-taxon discrimination, triplet metric learning~\citep{schroff2015facenet}, or direct taxonomic classification~\citep{MycoAI}, can provide an explicit taxonomy-related training signal to the global representation. Evidence from vision transformers further suggests that global-token design can substantially influence representation quality: register tokens~\citep{registers} reduce high-norm outliers that arise when patch tokens are reused for global aggregation, while Jumbo tokens~\citep{fuller2025jumbo} increase the capacity of the global token and improve performance over register-augmented baselines. Despite these advances, the roles of global-token design and supervision in DNA foundation models have not yet been systematically studied.

In this work, we introduce \textbf{BarcodeMAE+}, and use it to examine three design choices for DNA barcode foundation models: (1) \emph{model architecture}, comparing a standard encoder-only BERT model~\citep{devlin2019bert} with an encoder-decoder MAE-LM~\citep{meng2024maelm} architecture, which removes masked tokens from the encoder; (2) \emph{global representation}, examining whether the encoder benefits from an explicit \texttt{[CLS]} token; and (3) \emph{auxiliary pretraining objective}, comparing binary pairwise same-genus classification, triplet margin loss, and direct cross-entropy genus classification applied to the \texttt{[CLS]} representation. We evaluate these design choices across two distinct DNA barcode regions: arthropod COI using BIOSCAN-5M~\citep{gharaee2024bioscan5m}, and fungal ITS using UNITE+INSD~\citep{Abarenkov}, following the preprocessing of Romeijn et al.~\citep{MycoAI}. Our main contributions are:
\begin{itemize}
  \item We compare encoder-decoder MAE-LM pretraining with standard encoder-only MLM across ten model configurations and both barcode regions. We demonstrate that removing \texttt{[MASK]} tokens from the encoder consistently improves the quality of frozen representations in nearly all configurations, including the best configuration for each region.
  \item We investigate how global sequence representations should be learned by evaluating an explicit \texttt{[CLS]} token together with three auxiliary taxonomic objectives: binary pairwise same-genus classification, triplet margin loss, and direct cross-entropy genus classification. We demonstrate that auxiliary taxonomic objectives substantially improve the downstream performance of the \texttt{[CLS]} representation and that the choice of objective strongly influences the resulting representation. We further show that increasing the capacity of the global representation using a Jumbo \texttt{[CLS]} token does not improve performance over the standard \texttt{[CLS]} design.
  
  \item We examine the sensitivity of KNN evaluation to the neighbourhood size $k$ by comparing standard uniform voting with similarity-weighted softmax voting over an extended range of $ k$ values (from $ 1$ to $50$). We demonstrate that softmax weighting substantially stabilizes downstream accuracy across $k$, maintaining strong performance at large neighbourhood sizes where uniform-vote KNN degrades considerably.
\end{itemize}

\section{Method}
\label{sec:method}
\vspace{-4pt}
BarcodeMAE+ extends standard MLM pretraining with two complementary design components: (i) an encoder-decoder architecture in which the encoder processes only unmasked tokens, and (ii) an explicit global \texttt{[CLS]} representation trained with an auxiliary objective. The following subsections describe each component in detail.

\subsection{Datasets}
\vspace{-4pt}
\label{sec:datasets}
We evaluate our models on two barcode datasets comprising short, standardized marker
gene sequences: BIOSCAN-5M (arthropod COI) and UNITE+INSD (fungal ITS).
These two datasets differ substantially in taxonomic scope, label density, and
sequence diversity, allowing us to assess whether the investigated design choices
generalize across barcode regions and taxa.

\noindent\textbf{BIOSCAN-5M dataset.} BIOSCAN-5M~\citep{gharaee2024bioscan5m} is a large-scale collection of arthropod
Cytochrome c Oxidase Subunit I (COI) barcodes, a standardized 658\,bp mitochondrial
gene region widely used for specimen identification and species discovery.
The full dataset comprises 5.15 million arthropod specimens from which 2.4 million
unique DNA barcodes are derived.
We use 2.28 million sequences without species-level labels for unsupervised pretraining; genus-level annotations are available for 15.9\% of these sequences.
BIOSCAN-5M partitions its labelled data into a Seen split, containing species observed during pretraining, and an Unseen split, containing novel species entirely 
absent from pretraining. We use the training subset (118K) of the Seen split as the reference set against which test sequences are matched, and evaluate on the 
test subset (18.4K) of the Seen split and the test subset (3.4K) of the Unseen split; further details are provided in the Section \nameref{sec:Evaluation-Protocol}.
The taxonomic distribution of the KNN reference set and of the query
(test) set is provided in Supplementary Materials, Section A.

\noindent\textbf{UNITE+INSD.}
The UNITE+INSD dataset~\cite{Abarenkov}, preprocessed by~\citep{MycoAI},
contains 5.23 million sequences of the fungal Internal Transcribed Spacer (ITS) region, the standard
variable ribosomal marker for fungi, covering 14.7K distinct species organized across
seven taxonomic ranks: 1 kingdom (Fungi), 18 phyla, 70 classes, 231 orders, 791 families, 3,695 genera, 14.7k species.
The dataset presents a challenging partially-labelled setting: only 7\% of sequences
carry species-level annotations, while 80.3\% carry a genus-level annotation.
We pretrain on the full 5.23M corpus and evaluate on the two released held-out test sets: a yeast-specific benchmark (4,247 unique barcode sequences) and a filamentous fungal
set (10,873 unique barcode sequences). 
For our experiments, we exclude evaluation queries with any
sequence-level overlap with the training set, so the deduplicated query
pool used for evaluation is smaller than these released sizes; the exact deduplicated counts are given in the \nameref{s:experimental-setup}.
The taxonomic distribution of the KNN reference set and of the query
(test) sets is provided in Supplementary Materials, Section A.
\vspace{-10pt}
\subsection{Proposed Method}
\begin{figure*}[!t]
    \centering \includegraphics[width=\linewidth]{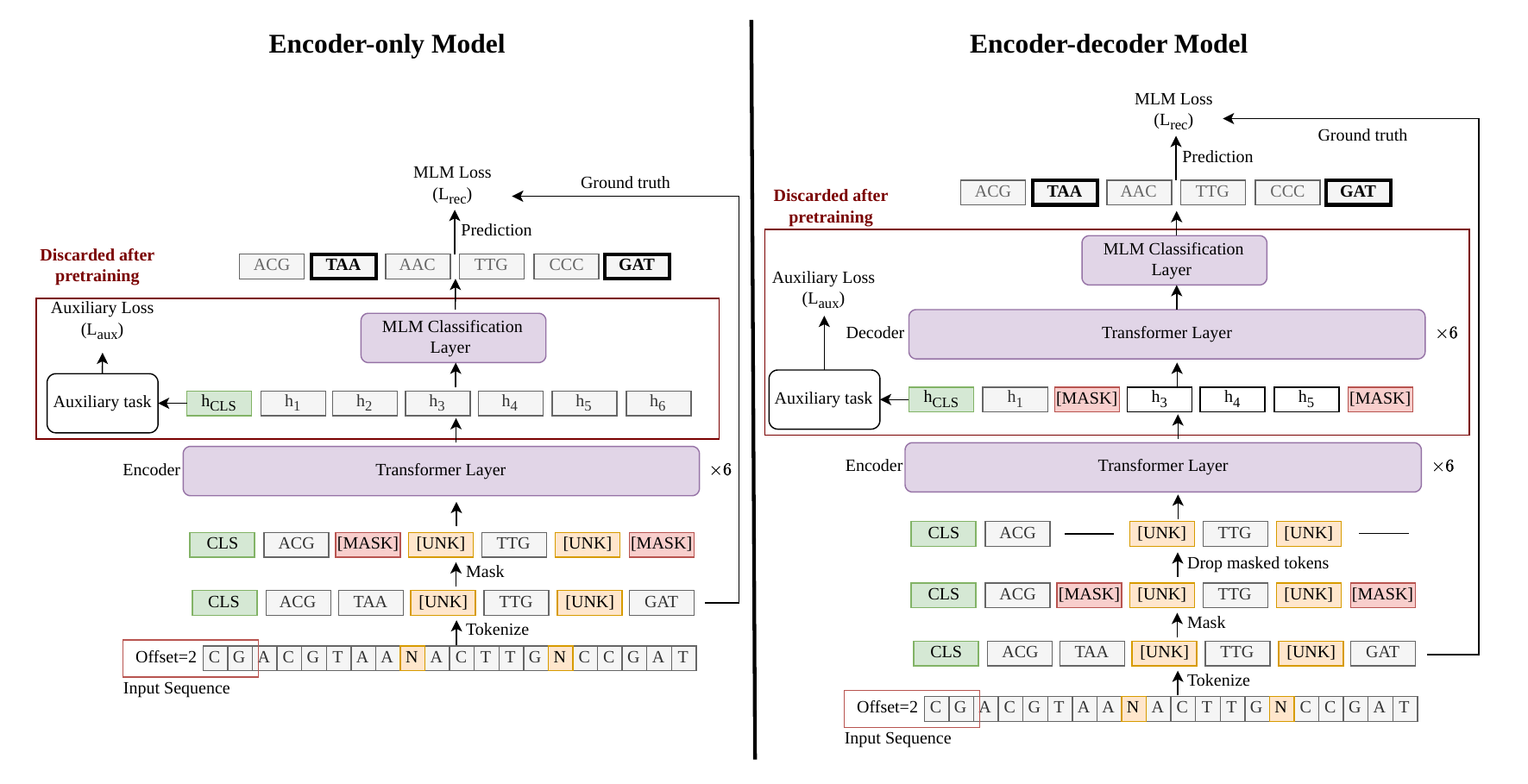}\caption{Architecture of the two encoder pretraining variants compared in this work, both operating on the same tokenized DNA sequence. Input sequences are tokenized into $k$-mers using a random offset $o \in \{0,\ldots,k-1\}$, improving robustness to mutations. All $k$-mers containing ambiguous nucleotides, \texttt{N}, are mapped to \texttt{[UNK]}. A random subset of tokens (50\%) is selected for masking, and an optional learnable \texttt{[CLS]} token is prepended, attending to all visible sequence tokens through standard multi-head self-attention. \textbf{Encoder-only model (left).} Masked tokens are replaced by \texttt{[MASK]} embeddings, and an encoder processes the full sequence. The encoder outputs are used directly to predict the masked tokens and, when applicable, to predict taxonomic labels from the \texttt{[CLS]} representation. \textbf{Encoder-decoder model (right).} Masked tokens are removed before the encoder, which therefore processes only the observed tokens. A decoder receives the encoder representations at observed positions and \texttt{[MASK]} embeddings at masked positions, and reconstructs the original tokens. When an auxiliary objective is used, the \texttt{[CLS]} representation from encoder is passed to a taxonomic head for pairwise binary classification, triplet learning, or cross-entropy classification. Only sequences with genus-level labels contribute to the auxiliary loss, $\mathcal{L}_{\mathrm{aux}}$, while all sequences contribute to $\mathcal{L}_{\mathrm{rec}}$. The total loss is $\mathcal{L}=\mathcal{L}_{\mathrm{rec}}+\alpha\mathcal{L}_{\mathrm{aux}}$. The decoder in the encoder-decoder model, and the auxiliary head in both models are discarded after pretraining, and only the encoder is used for downstream evaluation.}
    \label{fig:arch}
\end{figure*}

\vspace{-3pt}
\label{sec:maelm}
\textbf{Masked Autoencoder MLM.} To address the representational inefficiency in DNA sequence modelling, we adapt the MAE-LM approach \citep{meng2024maelm} for genomic applications. 
In training using masked language modelling objectives, part of the encoder's capacity must be allocated to processing \texttt{[MASK]} tokens, which potentially limits the model's overall representational capacity to encode real tokens. The MAE-LM architecture effectively mitigates this limitation by using a bidirectional encoder and bidirectional decoder, where the masked tokens are only presented to the decoder. \autoref{fig:arch} compares the MAE-LM approach (encoder-decoder architecture) against the encoder-only model.

As shown in \autoref{fig:arch}, the encoder-decoder's encoder operates on nucleotide sequences with masked-out tokens removed entirely. Given a DNA sequence $\mathbf{x} = [x_1, \ldots, [\texttt{MASK}]_i, \ldots, x_n]$ and the set of masked positions $\mathcal{M}$, the encoder processes only nucleotide tokens. The encoder's input sequence $\mathbf{H}^0$ is composed of token embeddings $\mathbf{e}_{x_i}$ and positional embeddings $\mathbf{p}_i$ for non-masked positions:
\begin{equation}
\mathbf{H}^0 = \{ \mathbf{h}_i^0\}_{i\notin\mathcal{M}}, \quad \mathbf{h}_i^0 = \mathbf{e}_{x_i} + \mathbf{p}_i
\end{equation}
The decoder then processes sequences containing both masked and unmasked positions, explicitly incorporating the \texttt{[MASK]} token in its input. The decoder's input sequence $\hat{\mathbf{H}}^0$ is constructed as:
\begin{equation}
\hat{\mathbf{H}}^0 = \{\hat{\mathbf{h}}_i^0\}_{1\leq i\leq n}, \quad
\hat{\mathbf{h}}_i^0 = \begin{cases}
\mathbf{e}_{[\text{MASK}]} + \mathbf{p}_i & i \in \mathcal{M} \\
\mathbf{h}_i^L + \mathbf{p}_i & i \notin \mathcal{M}
\end{cases}
\end{equation}
where $\mathbf{h}_i^L$ represents the final layer output from the encoder for non-masked positions and $\mathbf{e}_{[\text{MASK}]}$ is the token embedding for the \texttt{[MASK]} token.

This approach prevents the encoder from learning specific embeddings for the \texttt{[MASK]} token, ensuring the encoder's representational capacity is not devoted to encoding this special token. Consequently, the encoder's representations remain unaffected by the \texttt{[MASK]} token and will use the full representational capacity to learn meaningful patterns from the nucleotide sequences. During downstream tasks, only the encoder is used, effectively isolating potential limitations or inefficiencies related to the \texttt{[MASK]} tokens.

Note that, in this model, the encoder processes DNA sequences without \texttt{[MASK]} tokens, requiring a modified
positional encoding scheme. Our implementation preserves sequence order by skipping masked
position indices during encoding. This design maintains the relative positions of unmasked tokens
from the original sequence, enabling spatial relationship modelling in DNA sequences.

\noindent\textbf{Global Representation Token.} \label{sec:cls} Beyond token-level representations, barcode classification can benefit from a compact embedding that summarizes information across the full sequence. To obtain this global representation, we prepend a learnable \texttt{[CLS]} token to the encoder input, allowing it to attend jointly to all visible DNA tokens through standard multi-head self-attention. At inference, the final-layer output of the \texttt{[CLS]} token,
$\mathbf{h}^L_{\texttt{CLS}} \!\in\! \mathbb{R}^{D}$, serves as the global sequence embedding.
We also evaluate the Jumbo CLS token~\citep{fuller2025jumbo} as an enhanced alternative.
Rather than a single $D$-dimensional token, the Jumbo CLS token is $J$ times wider,
$\mathbf{h}^L_{\text{JumboCLS}} \!\in\! \mathbb{R}^{J \cdot D}$, and is processed after each attention
layer by its own dedicated wide feed-forward network with parameters shared across encoder layers.

\noindent\textbf{Auxiliary Taxonomic Objectives.}
\label{sec:aux}
The \texttt{[CLS]} token provides a dedicated global representation, but the MLM objective does not directly optimize this representation, as the reconstruction loss is defined over masked token predictions. To provide an additional training signal for the global representation, we attach an auxiliary objective to the \texttt{[CLS]} token during pretraining.  We supervise this objective at the genus rank: it is the finest rank densely labelled in both datasets, and it matches the rank used for downstream evaluation. Coarser (family) and finer (BIN) label supervision are evaluated in Supplementary Materials, Section B, and did not improve performance. The auxiliary head is discarded after pretraining. This auxiliary objective is applied only to sequences with genus-level taxonomic labels (including those identified to species), while all sequences continue to contribute to the reconstruction objective. We evaluate three auxiliary objective variants:

\noindent\textit{Binary same-genus prediction.}
For each pair of DNA sequences $(a, b)$, the head predicts whether they belong to the same
genus by concatenating their CLS representations and passing them through an MLP head:
\begin{equation}
  \hat{y}_{ab} = \sigma\!\bigl(
    f_\phi\bigl([
      \mathbf{h}_{\texttt{CLS}}^{(a)}
      \;\|\;
      \mathbf{h}_{\texttt{CLS}}^{(b)}
    ]\bigr)\bigr),
\end{equation}
where $f_\phi : \mathbb{R}^{2D} \to \mathbb{R}$ is an MLP
with hidden dimension 256, $\|$ denotes concatenation,
and $\sigma$ is the sigmoid function.
Pairs are sampled within each mini-batch, balanced between positive (same genus) and
negative (different genus) pairs.

\noindent\textit{Triplet margin loss.}
Given an anchor sequence $a$, a positive example $p$ (same genus as $a$), and a negative
example $n$ (different genus), we optimize:
\begin{equation}
  \mathcal{L}_{\mathrm{triplet}} = \max\!\bigl(0,\;
    d(\mathbf{h}_{\texttt{CLS}}^{a},\, \mathbf{h}_{\texttt{CLS}}^{p})
    - d(\mathbf{h}_{\texttt{CLS}}^{a},\, \mathbf{h}_{\texttt{CLS}}^{n}) + m
  \bigr),
\end{equation}
where $d(\cdot,\cdot)$ is cosine distance and $m$ is the margin hyperparameter.
Triplets are mined online within each mini-batch. We evaluate two triplet mining strategies within each mini-batch: batch-hard mining and random mining. In batch-hard mining, for each anchor we select the hardest positive, defined as the same-genus example with the largest distance from the anchor, and the hardest negative, defined as the different-genus example with the smallest distance from the anchor. In random mining, the positive and negative examples are sampled randomly from the same-genus and different-genus candidates within the mini-batch, respectively.

\noindent\textit{Direct genus classification (CE).} A linear classification head maps the \texttt{[CLS]} representation to a probability distribution over the C observed genera using a softmax function, and is optimized with the standard cross-entropy loss.

\noindent\textbf{Full Training Objective.} The full training objective combines the MAE-LM reconstruction loss with the chosen auxiliary loss:
\begin{equation}
  \mathcal{L} = \mathcal{L}_{\text{rec}} + \alpha\,\mathcal{L}_{\text{aux}}, \label{eq:loss}
\end{equation}
where $\mathcal{L}_{\text{aux}}$ is one of the three auxiliary losses described above
(binary cross-entropy, triplet, or CE), and $\alpha$ controls the relative weight.
Since the auxiliary objectives are formed only from sequences carrying genus-level labels,
the remaining sequences contribute only to $\mathcal{L}_{\text{rec}}$.

\noindent\textbf{Model Variants.}
For BIOSCAN-5M and UNITE+INSD, we train configurations spanning two encoder architectures (MAE-LM encoder-decoder and standard encoder-only Transformer), the presence or absence of a \texttt{[CLS]} token, and the choice of auxiliary objective. The resulting ten configurations are summarized in \autoref{tab:variants}. We also include a randomly initialized, untrained encoder as a baseline.

\begin{table}[t]
\centering
\caption{Pretraining configurations evaluated on the barcode datasets (BIOSCAN-5M and UNITE+INSD).
The Architecture column indicates whether the architecture uses an encoder-decoder MAE-LM
or an encoder-only model. The CLS column indicates whether a global CLS token
is prepended to the encoder input. The Aux column indicates the auxiliary
taxonomic objective attached to the \texttt{[CLS]} token; ``--'' means reconstruction loss only.}
\label{tab:variants}
\small
\begin{tabular}{clcc}
\toprule
\# & \textbf{Architecture} & \textbf{CLS} & \textbf{Aux} \\
\midrule
1 & Encoder-decoder          & \texttimes & -- \\
2 & Encoder-decoder           & \checkmark & -- \\
3 & Encoder-decoder           & \checkmark & Binary \\
4 & Encoder-decoder           & \checkmark & Triplet \\
5 & Encoder-decoder           & \checkmark & CE \\
\midrule
6 & Encoder-only      & \texttimes & -- \\
7 & Encoder-only      & \checkmark & -- \\
8 & Encoder-only      & \checkmark & Binary \\
9 & Encoder-only      & \checkmark & Triplet \\
10& Encoder-only      & \checkmark & CE \\
\bottomrule
\end{tabular}
\end{table}
\section{Experimental Setup}
\vspace{-3pt}
\label{s:experimental-setup}
This section describes the experimental configuration used to pretrain and evaluate the proposed models.
\vspace{-3pt}
\subsection{Experimental Configuration}
\vspace{-3pt}
\textbf{Architecture.}
All encoder-decoder models use a 6-layer, 6-head Transformer encoder with hidden dimension $D=768$, following the BarcodeMAE architecture~\citep{safari2025barcodeMAE}. During pretraining, these models include a matching 6-layer, 6-head decoder, which is discarded after pretraining so that only the encoder is used at inference. The encoder-only baseline uses the same encoder configuration, ensuring that both architectures have identical encoder capacity at inference and differ only in their pretraining design. \autoref{fig:arch} compares the full architecture of both variants.

\noindent\textbf{Tokenization.}
For the BIOSCAN-5M and UNITE+INSD datasets, we adopt $k$-mer tokenization, motivated by the finding in BarcodeBERT~\citep{millan2023barcodebert} that $k$-mer tokenization substantially outperforms Byte Pair Encoding (BPE)~\citep{sennrich2016bpe} on barcode data. Sequences are split into non-overlapping $k$-mers with $k=6$, following the BarcodeBERT study~\citep{millan2023barcodebert}, yielding a vocabulary of $4^6 + 3 = 4099$ tokens, including the three special tokens \texttt{[CLS]}, \texttt{[MASK]}, and \texttt{[UNK]}. Any $k$-mer containing an ambiguous nucleotide is mapped to \texttt{[UNK]}. A key limitation of $k$-mer tokenization is its sensitivity to frame shifts: offsetting a sequence by a single nucleotide produces an entirely different token sequence. Following BarcodeBERT~\citep{millan2023barcodebert}, we mitigate this during pretraining by randomly shifting each sequence by $r \sim \mathrm{Uniform}\{0,1,\ldots,k-1\}$ nucleotides before tokenization.

\noindent\textbf{Pretraining Hyperparameters.}
We implement all models in PyTorch using the Hugging Face Transformers library. For both BIOSCAN-5M and UNITE+INSD, we train with a masked token prediction objective using a 50\% masking ratio, following BarcodeBERT \citep{millan2023barcodebert}. The cross-entropy loss over masked tokens is optimized with AdamW~\citep{loshchilov2017adamw} with weight decay $1\times10^{-5}$ and a OneCycle~\cite{OneCycle} learning rate schedule with a maximum learning rate of $7\times10^{-5}$. We train in BF16 mixed precision with a batch size of 128. For the auxiliary pairwise classification task, each batch contains a balanced set of up to 64 positive and 64 negative pairs, for a total of up to 128 pairs. We set the auxiliary-loss weight to $\alpha=1.0$ for BIOSCAN-5M and $\alpha=0.1$ for UNITE+INSD. For each dataset, we selected the value that performed best with the chosen auxiliary objective from $\alpha \in \{0.01, 0.05, 0.1, 0.5, 1.0\}$. Ideally, this hyperparameter would be selected using a held-out validation set. However, our evaluation setting requires genus-level $k$-nearest-neighbour (KNN) classification of unseen species, and neither BIOSCAN-5M nor UNITE+INSD, externally curated benchmarks that we did not construct, provides a validation split satisfying this condition. We therefore checked $\alpha$ directly on performance on the evaluation sets.

\noindent\textbf{Sequence Representations.}
After pretraining, the encoder is used to extract sequence embeddings under several representation strategies, summarized in \autoref{tab:representations}. For models with a \texttt{[CLS]} token, we evaluate three strategies. \textbf{Tokens} averages the encoder outputs over all DNA sequence tokens, excluding \texttt{[CLS]}. \textbf{CLS} uses the output of the \texttt{[CLS]} token as the global sequence embedding. \textbf{Tokens+CLS} averages over all encoder outputs, including both the DNA tokens and \texttt{[CLS]}. For models without a \texttt{[CLS]} token, only \textbf{Tokens} is applicable.

\begin{table}[t]
\centering
\caption{Sequence representation strategies used for evaluation in this work.}
\label{tab:representations}
\small
\begin{tabular}{ll}
\toprule
\textbf{Representation} & \textbf{Description} \\
\midrule
Tokens        & Mean over DNA tokens only (excluding CLS) \\
CLS           & Output of \texttt{[CLS]} token \\
Tokens+CLS    & Mean over all encoder outputs (DNA tokens + CLS) \\
\bottomrule
\end{tabular}
\end{table}
\vspace{-3pt}
\subsection{Evaluation Protocol}\label{sec:Evaluation-Protocol}
\vspace{-3pt}
\textbf{KNN Evaluation.}
For both barcode datasets, we perform KNN probing on frozen encoder embeddings using cosine similarity, without fine-tuning the encoder. We evaluate two strategies for predicting the class of each query from its $k$ nearest neighbours: uniform voting and similarity-weighted softmax voting.

\textit{Uniform voting.}
Each of the $k$ nearest neighbours casts an equal vote for its class label, and the predicted class is the majority class among the $k$ neighbours.

\textit{Softmax voting.}
Following the similarity-weighted KNN protocol of DINOv2~\citep{oquab2024dinov2}, each neighbour $i$ with cosine similarity $s_i$ to the query and class label $y_i$ contributes a temperature-scaled softmax weight,
\begin{equation}
  w_i = \frac{\exp(s_i / T)}{\sum_{j=1}^{k} \exp(s_j / T)}.
\end{equation}
\noindent where $T>0$ is the temperature.
The predicted class is given by $\arg\max_c \sum_{i:\, y_i = c} w_i$, the class receiving the largest total weight among the $k$ neighbours. The temperature $T$ controls how sharply influence decays with similarity: as $T \to 0$ all weight concentrates on the single nearest neighbour with the rule reducing to 1-NN regardless of $k$, while as $T$ grows the weights flatten and the rule approaches uniform voting. Empirically we found $T=0.02$ works well and adopt it as the default temperature for our experiments. At this value, there is a sharp dropoff in influence for neighbours away from the closest neighbour: a neighbour $0.1$ less similar than the closest receives roughly $1/150$ of its weight. Results for different values of $T$, $T \in \{0.01,0.02,0.05,0.07,0.1,0.2,0.5,1.0\}$ are reported in Supplementary Materials, Section~C. We evaluate both voting strategies across $k \in \{1,3,5,7,10,15,20,25,50\}$. For the baseline comparisons in ~\autoref{tab:bioscan_external} and~\autoref{tab:its_external}, we report each model's best softmax accuracy across the full $k \times T$ grid, ensuring a fair comparison across all baselines; the corresponding best $(T,k)$ pairs are reported in Supplementary Materials, Section~D.

\noindent\textbf{BIOSCAN-5M.}
We evaluate BIOSCAN-5M using two complementary tasks: genus-level KNN classification and zero-shot BIN reconstruction.

\noindent\textit{Genus-level KNN classification.}
We use the training subset of the \textit{Seen} partition as the reference set and the test subset of the \textit{Unseen} partition as the query set. Although the query samples belong to species that were not observed during training, each of their genera is represented in the reference set. Accordingly, this task operates within a closed-world setting and evaluates the model's ability to generalize to new species within known genera. We report genus-level accuracy as the primary metric. 

\noindent\textit{Zero-shot BIN reconstruction.}
To assess the model's ability to identify novel species and capture taxonomic relationships, we implement a Barcode Index Number (BIN) reconstruction task. We merge the test subset from the \textit{Seen} partition with the test subset of the \textit{Unseen} partition and perform zero-shot clustering on embeddings generated without fine-tuning, following~\citep{zsc-Lowe-2024}. Embeddings are reduced to 50 dimensions using UMAP~\citep{mcinnes2018umap} and clustered using agglomerative clustering using euclidean distance with Ward's linkage, with the number of BINs used as the target number of clusters. Clustering quality is measured using Adjusted Mutual Information (AMI) against ground-truth labels.

\noindent\textbf{UNITE+INSD.}
We evaluate UNITE+INSD using genus-level KNN classification on two held-out fungal ITS test sets.

\noindent\textit{Genus-level KNN classification.}
The 5.23M-sequence training set is used as the KNN reference set. We evaluate on two held-out query sets: \textit{Yeast}~\citep{vu2016yeast} and \textit{Filamentous Fungi}~\citep{vu2019filamentous}. We exclude any query that overlaps with the training set, based on both the sample ID and the exact barcode sequence. For Yeast, this removes 3,721 sequences, leaving 526 genus-level test sequences. For Filamentous Fungi, this removes 7,737 sequences, leaving 3,136 genus-level test sequences (full breakdown in Supplementary Materials, Section A). We report genus-level accuracy as the primary metric.
\vspace{-10pt}
\section{Results}
\label{sec:results}
\vspace{-3pt}
We compare BarcodeMAE+ with published DNA foundation models using frozen encoder representations. After selecting the best configuration (encoder-decoder MAE-LM with a trained \texttt{[CLS]} token and dataset-appropriate auxiliary objective: CE on BIOSCAN-5M, Binary on UNITE+INSD), BarcodeMAE+ outperforms the evaluated baselines on BIOSCAN-5M in both genus-level KNN accuracy and zero-shot BIN reconstruction AMI, and achieves the highest Yeast accuracy on UNITE+INSD. \autoref{tab:bioscan_external} compares our best configuration against other DNA foundation models evaluated on BIOSCAN-5M. We evaluate against models pretrained on a range of corpora, including the human genome, broader multi-species assemblies, Canadian invertebrate COI barcodes (Canada-1.5M)~\citep{dewaard2019reference}, and BIOSCAN-5M itself. BarcodeMAE+ achieves the best result on both tasks: genus-level KNN accuracy and zero-shot BIN reconstruction AMI. \autoref{tab:its_external} compares our best UNITE+INSD configuration with other DNA foundation models and fungal ITS-specific baselines on the deduplicated query sets. BarcodeMAE+ achieves the highest Yeast accuracy at 73.19\%, outperforming the strongest external baseline, MycoAI-BERT, at 70.48\%. On Filamentous, BarcodeMAE+ reaches 63.07\%, slightly below MycoAI-BERT at 64.16\% and above MycoAI-CNN at 62.31\%. Unlike BarcodeMAE+, both MycoAI models are trained end-to-end using taxonomic labels, whereas BarcodeMAE+ is evaluated using frozen encoder representations. Note that \autoref{tab:its_external} reports results on the deduplicated UNITE+INSD query pools; results on the original published test sets are provided in Supplementary Materials, Section E.
\vspace{-5pt}
\begin{table*}[!h]
\centering
\footnotesize
\caption{Comparison with DNA foundation model baselines, evaluated zero-shot
(frozen embeddings, no fine-tuning) on the unseen-species test partition of \textbf{BIOSCAN-5M}. Baselines are pretrained on a range of
corpora. We report genus-level 1-NN accuracy ($k{=}1$, closed-world), the best softmax-voted
KNN accuracy minus the 1-NN accuracy, in percentage points (Softmax $\Delta$), searched over $k \in \{1,3,5,7,10,15,20,25,50\}$ and $T \in \{0.01,0.02,0.05,0.07,0.1,0.2,0.5,1.0\}$,
zero-shot BIN reconstruction AMI (open-world), and their harmonic mean (HM).
A value of \textcolor{gray}{0.00} (shown in gray) indicates that the best softmax accuracy occurs at $k=1$, where softmax weighting reduces to 1-NN voting, so the value is identical to the 1-NN column. None of these models are trained end-to-end via full supervised
training; all are evaluated on frozen embeddings. Training task abbreviations are, NTP = next-token
prediction, MLM = masked language modelling, CL = contrastive learning, MLM+CE = masked language
modelling with a cross-entropy genus-classification auxiliary objective. The best value is shown in \textbf{bold},
and the second-best value is \underline{underlined}.}
\label{tab:bioscan_external}
\setlength{\tabcolsep}{5pt}
\begin{tabular}{llllcccc}
\toprule
Architecture & Pretraining data & Model & Training task & 1-NN Acc. & Softmax $\Delta$ (pp) & BIN AMI & HM \\
\midrule
\multirow{4}{*}{State space}
  & Human genome      & HyenaDNA-tiny~\cite{nguyen2024hyenadna}        & NTP & 20.52 & \textcolor{gray}{0.00} & 61.26 & 30.74 \\
  & Human genome      & Caduceus-PS-1k~\cite{schiff2024caduceus}       & MLM & 12.84 & \textcolor{gray}{0.00} & 54.92 & 20.81 \\
  & BIOSCAN-5M        & BarcodeMamba+~\cite{gao2024barcodemamba}        & NTP & 29.47 & \textcolor{gray}{0.00} & 66.78 & 40.89 \\
\midrule
\multirow{6}{*}{Encoder-only}
  & Multi-species DNA & DNABERT-2~\cite{zhou2023dnabert2}              & MLM & 16.08 & \textcolor{gray}{0.00} & 30.70 & 21.11  \\
  & Multi-species DNA & DNABERT-S~\cite{zhou2024dnabertS}              & CL  & 19.41 & +1.17 (k=20) & 60.25 & 29.36 \\
  & Multi-species DNA & Nucleotide Transformer~\cite{dalla2024nucleotide} & MLM & 22.88 & \textcolor{gray}{0.00} & 32.52 & 26.86 \\
  & Human genome      & GENA-LM~\cite{fishman2025genalm}               & MLM & 24.91 & \textcolor{gray}{0.00} & 60.25 & 35.25 \\
  & Canada-1.5M       & BarcodeBERT~\cite{millan2023barcodebert}       & MLM & 40.96 & \textcolor{gray}{0.00} & 68.36 & 51.23 \\
  & BIOSCAN-5M        & BarcodeBERT~\cite{millan2023barcodebert}       & MLM & \underline{58.31}  & \textcolor{gray}{0.00} & \underline{72.58} & \underline{64.67} \\
\midrule
Encoder-decoder
  & BIOSCAN-5M        & \textbf{BarcodeMAE+ (ours)}                    & MLM+CE & \textbf{80.65} & \textcolor{gray}{0.00} & \textbf{75.79} & \textbf{78.15} \\
\bottomrule
\end{tabular}
\end{table*}

\begin{table*}[!h]
\centering
\footnotesize
\caption{Comparison with baselines on \textbf{UNITE+INSD}, evaluated
using frozen embeddings on the deduplicated Yeast and Filamentous query
pools. We report genus-level 1-NN accuracy and the
best softmax-voted KNN accuracy minus the 1-NN accuracy, in percentage points
(Softmax $\Delta$), searched over
$k \in \{1,3,5,7,10,15,20,25,50\}$ and
$T \in \{0.01,0.02,0.05,0.07,0.1,0.2,0.5,1.0\}$ for each test set.
A value of \textcolor{gray}{0.00} (shown in gray) indicates that the best softmax
accuracy occurs at $k=1$, where softmax weighting reduces to 1-NN voting, so
the value is identical to the 1-NN column. Training task abbreviations are, NTP = next-token prediction,
MLM = masked language modelling, CL = contrastive learning,
SC = supervised classification, MLM+Binary = masked language modelling with a same-genus
pairwise auxiliary objective. The best
value in each 1-NN column is shown in \textbf{bold}, and the second-best
value is \underline{underlined}.}
\label{tab:its_external}

\setlength{\tabcolsep}{4pt}
\begin{tabular}{lllccccc}
\toprule
Architecture
& Training data
& Model
& Training task
& \multicolumn{2}{c}{Yeast}
& \multicolumn{2}{c}{Filamentous} \\
\cmidrule(lr){5-6}
\cmidrule(lr){7-8}
&
&
&
& 1-NN
& Softmax $\Delta$ (pp)
& 1-NN
& Softmax $\Delta$ (pp) \\
\midrule

\multirow{4}{*}{State space}
  & Human genome
  & HyenaDNA-tiny~\cite{nguyen2024hyenadna}
  & NTP
  & 35.17
  & \textcolor{gray}{0.00}
  & 37.72
  & \textcolor{gray}{0.00} \\

  & Human genome
  & Caduceus-PS-1k~\cite{schiff2024caduceus}
  & MLM
  & 27.95
  & \textcolor{gray}{0.00}
  & 33.51
  & \textcolor{gray}{0.00} \\

  & UNITE+INSD
  & BarcodeMamba+~\cite{gao2025barcodemambaplus}
  & NTP
  & 62.35
  & \textcolor{gray}{0.00}
  & 55.58
  & +1.46 (k=5) \\

  & UNITE+INSD
  & BarcodeMamba+ (large)~\cite{gao2025barcodemambaplus}
  & NTP
  & 61.03
  & +1.51 (k=7)
  & 55.49
  & +0.82 (k=5) \\

\midrule

\multirow{6}{*}{Encoder-only}
  & Multi-species DNA
  & DNABERT-2~\cite{zhou2023dnabert2}
  & MLM
  & 61.03
  & \textcolor{gray}{0.00}
  & 54.02
  & +0.13 (k=3) \\

  & Multi-species DNA
  & DNABERT-S~\cite{zhou2024dnabertS}
  & CL
  & 50.38
  & +0.57 (k=5)
  & 51.56
  & +0.61 (k=5) \\

  & Multi-species DNA
  & Nucleotide Transformer~\cite{dalla2024nucleotide}
  & MLM
  & 36.69
  & \textcolor{gray}{0.00}
  & 41.87
  & \textcolor{gray}{0.00} \\

  & Canada-1.5M
  & BarcodeBERT~\cite{millan2023barcodebert}
  & MLM
  & 18.82
  & \textcolor{gray}{0.00}
  & 32.02
  & \textcolor{gray}{0.00} \\

  & Human genome
  & GENA-LM~\cite{fishman2025genalm}
  & MLM
  & 55.32
  & \textcolor{gray}{0.00}
  & 55.90
  & \textcolor{gray}{0.00} \\

  & UNITE+INSD
  & MycoAI-BERT~\cite{MycoAI}
  & SC
  & \underline{70.48}
  & +2.33 (k=5)
  & \textbf{64.16}
  & +1.85 (k=15) \\

\midrule

CNN
  & UNITE+INSD
  & MycoAI-CNN~\cite{MycoAI}
  & SC
  & 69.01
  & +2.66 (k=5)
  & 62.31
  & +1.62 (k=5) \\

\midrule

Encoder-decoder
  & UNITE+INSD
  & \textbf{BarcodeMAE+ (Ours)}
  & MLM+Binary
  & \textbf{73.19}
  & +3.24 (k=25)
  & \underline{63.07}
  & +2.14 (k=50) \\

\bottomrule
\end{tabular}
\end{table*}
\vspace{-3pt}
\subsection{Encoder-decoder vs. encoder-only architecture}
\begin{figure*}[t]
\centering
\includegraphics[width=\textwidth]{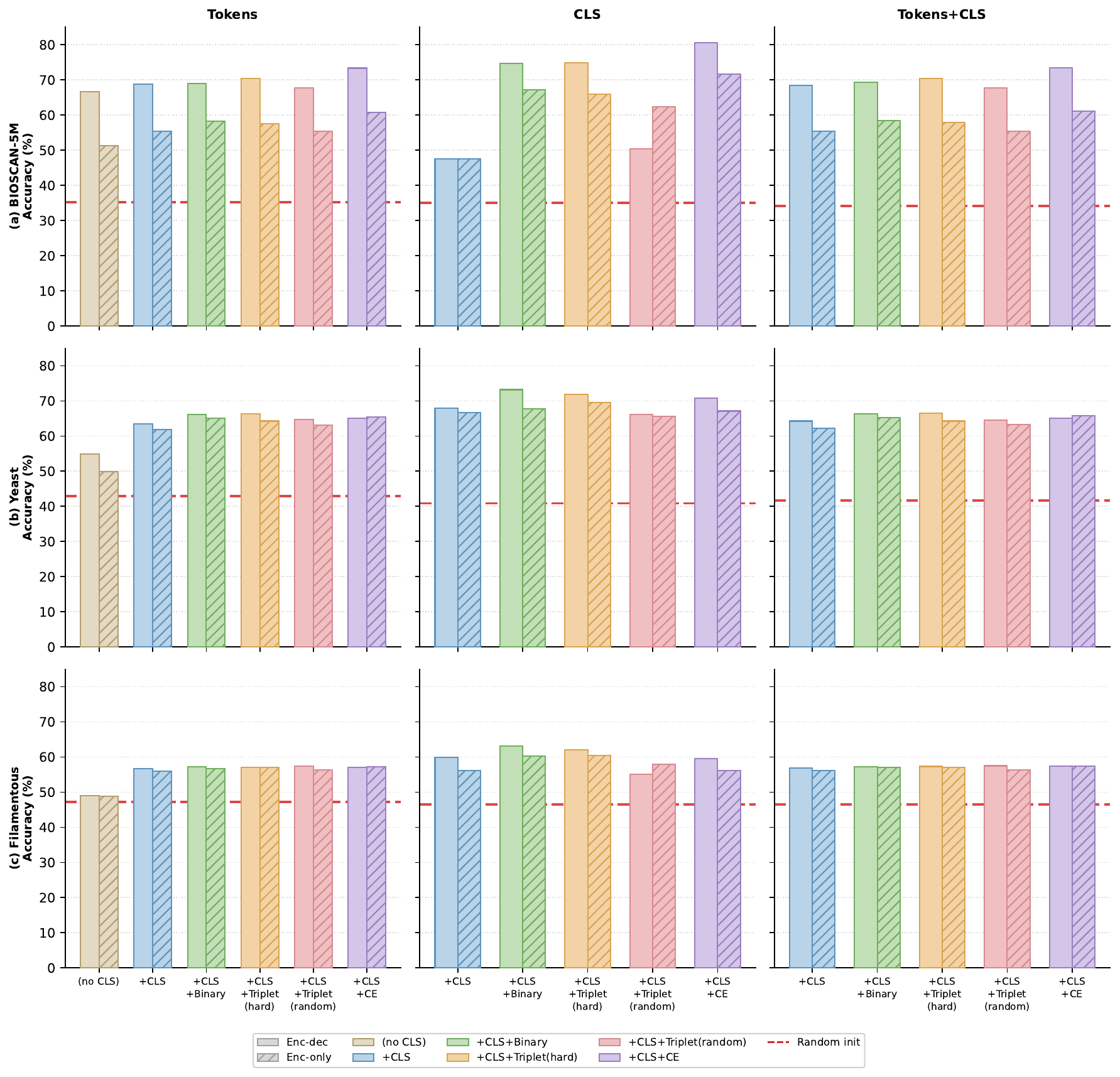}
\caption{Genus-level 1-NN ($k{=}1$) accuracy across all architectures, auxiliary
objectives, and sequence representations (Tokens, CLS, Tokens+CLS columns), for (a)
BIOSCAN-5M, (b) UNITE+INSD Yeast, and (c) UNITE+INSD Filamentous. The red dash-dot line
shows the randomly-initialized encoder baseline.}
\label{fig:combined_bar_k1}
\end{figure*}

\vspace{-3pt}
\label{sec:results:bioscan}
Across both datasets, encoder–decoder MAE-LM pretraining yields the best-performing configuration and outperforms the matched encoder-only baseline in nearly all downstream \(k=1\) comparisons. On BIOSCAN-5M, the improvement is most pronounced for token-based representations, with gains of 11--15 percentage points for Tokens and Tokens+CLS; for example, the no-CLS configuration improves from 51.27\% to 66.73\%. When a global \texttt{[CLS]} token is present, the encoder-decoder advantage persists but is smaller (7.5--9.0 percentage points across CE, Binary, and batch-hard Triplet) and is amplified when a CLS-specific auxiliary objective is used. \autoref{fig:combined_bar_k1}--a reports genus-level $k{=}1$ KNN accuracy across all architectures, auxiliary objectives, and sequence representations, including a randomly initialized encoder baseline.

In the BIN reconstruction task, ~\autoref{tab:zsc_results} reports zero-shot BIN reconstruction performance using Adjusted Mutual Information (AMI). The encoder-decoder architecture again outperforms the encoder-only baseline across all configurations. The largest gap appears without a \texttt{[CLS]} token, where the Tokens representation achieves 74.32\% AMI compared with 55.65\% for the encoder-only model. This difference becomes much smaller when an auxiliary objective is used with the \texttt{[CLS]} representation; for example, the CE configurations achieve 75.79\% and 74.45\% AMI for the encoder-decoder and encoder-only models, respectively. Unlike the genus-level KNN results, the choice of auxiliary objective has little effect on zero-shot clustering performance, with the encoder-decoder \texttt{[CLS]} configurations ranging only from  74.31\% to 75.79\% AMI. Additional zero-shot clustering results across all model configurations and sequence representation strategies are reported in Supplementary Materials, Section~F.

Unlike BIOSCAN-5M, where the encoder-decoder architecture provides its largest improvement over the encoder-only baseline for token-based representations without a \texttt{[CLS]} token, the pattern is different on UNITE+INSD. As shown in \autoref{fig:combined_bar_k1}--b and \autoref{fig:combined_bar_k1}--c, the advantage of the encoder-decoder architecture is more pronounced for the CLS representation when an auxiliary objective is used. Without a \texttt{[CLS]} token, the encoder-decoder model improves the Tokens representation by 5.13 percentage points on Yeast (54.94\% vs.\ 49.81\%) and only 0.26 percentage points on Filamentous (49.11\% vs.\ 48.85\%). With an auxiliary objective, the encoder-decoder architecture improves the CLS representation by up to 5.5 percentage points on Yeast and 3.35 percentage points on Filamentous.

\begin{table}[!h]
\centering
\scriptsize
\caption{Zero-shot BIN reconstruction AMI (\%) on BIOSCAN-5M for all five
configurations, comparing the encoder-decoder and encoder-only architectures. The best value is shown in \textbf{bold}, and the second-best value is \underline{underlined}.}
\label{tab:zsc_results}
\setlength{\tabcolsep}{5pt}
\begin{tabular}{llcc}
\toprule
 & & Encoder-decoder & Encoder-only \\
\cmidrule(lr){3-3} \cmidrule(lr){4-4}
Configuration & Representation & BIN AMI & BIN AMI \\
\midrule
Base (no CLS)& Tokens & 74.32 & 55.65 \\
+CLS         & CLS    & 54.94 & 51.78 \\
+CLS+Binary  & CLS    & \underline{75.02} & 73.39 \\
+CLS+Triplet & CLS    & 74.31 & 72.76 \\
+CLS+CE      & CLS    & \textbf{75.79} & 74.45 \\
\bottomrule
\end{tabular}
\end{table}

\vspace{-3pt}
\subsection{Auxiliary objectives for the global representation}

\autoref{fig:combined_bar_k1} reports genus-level 1-NN accuracy for every configuration, representation, and dataset; the values quoted in this section are read from that figure, with the underlying numbers in Supplementary Materials, Section~G.
We first note that the CLS representation performs poorly without an auxiliary objective, reaching only 47.53\% accuracy at $k{=}1$ on BIOSCAN-5M, and 67.87\% (Yeast) / 59.95\% (Filamentous) on UNITE+INSD. Adding an auxiliary taxonomic objective substantially improves the global representation, with the best objective being dataset-dependent: on BIOSCAN-5M, CE yields the highest accuracy (80.65\%), followed by Triplet with batch-hard mining (74.85\%) and Binary (74.76\%); on UNITE+INSD, Binary same-genus prediction attains the highest accuracies (Yeast 73.19\%, Filamentous 63.07\%); \autoref{tab:its_external}. For Triplet, the mining strategy matters for the CLS representation: batch-hard preserves performance (74.85\%) while random mining drops to 50.32\%. The Tokens representation is less sensitive (70.44\% vs.\ 67.76\%), suggesting the learned global \texttt{[CLS]} depends more than token averaging on a strong auxiliary signal. Notably, under random mining the encoder-only baseline outperforms encoder-decoder for the CLS representation on some, but not all, test sets, though neither surpasses the dataset's best configuration. Changing the auxiliary labels (family or BIN) or replacing the standard \texttt{[CLS]} with a wider Jumbo token did not improve over the dataset-specific best configuration on either BIOSCAN-5M or UNITE+INSD (best family/BIN: 72.61\% BIOSCAN-5M, 71.67\%/60.52\% UNITE+INSD Yeast/Filamentous; best Jumbo: 69.38\% BIOSCAN-5M, 65.02\%/56.25\% UNITE+INSD Yeast/Filamentous); see Supplementary Materials, Sections~B and~H for details. This dataset-dependent pattern motivated the selection of CE for BIOSCAN-5M and Binary for UNITE+INSD in our reported comparisons.
\vspace{-5pt}
\subsection{Softmax vs.\ uniform voting stability.}
\vspace{-3pt}
\begin{figure*}[htp]
\centering
\includegraphics[width=\textwidth]{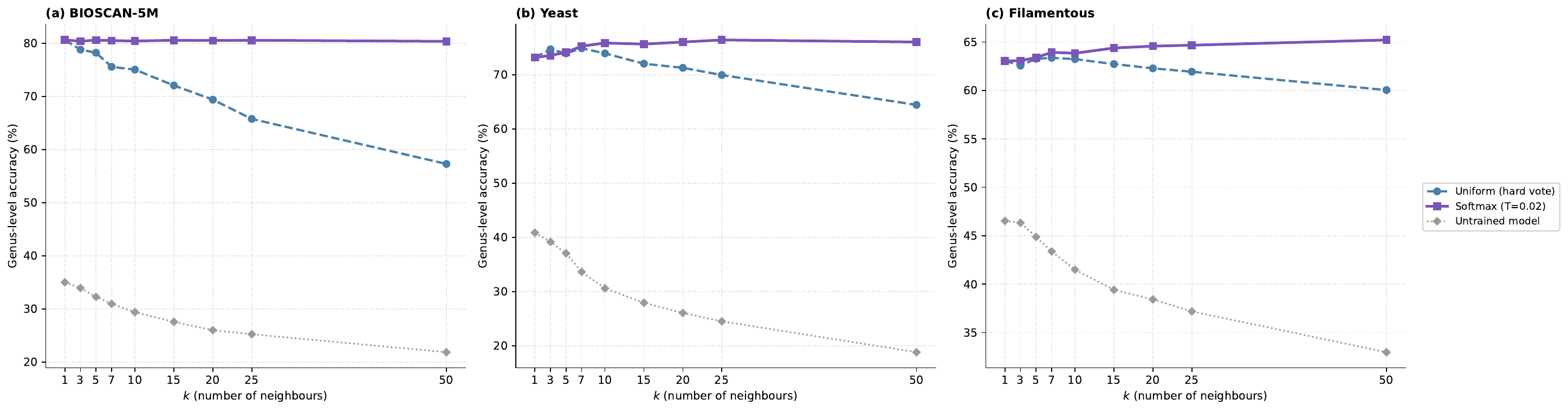}
\caption{Genus-level KNN accuracy vs.\ $k$ for each dataset's best configuration:
(a) BIOSCAN-5M (encoder-decoder, +CLS+CE, CLS representation), (b) UNITE+INSD Yeast
and (c) UNITE+INSD Filamentous (encoder-decoder, +CLS+Binary, CLS representation),
comparing uniform voting (dashed), softmax voting (solid, $T=0.02$), and an untrained
model (dotted). Softmax voting substantially stabilizes accuracy at large $k$ in all
cases.}
\label{fig:main_kcurve}
\end{figure*}

For BIOSCAN-5M, the best-performing configuration is highly sensitive to the choice of $k$ under uniform voting, with accuracy decreasing from 80.65\% at $k{=}1$ to 57.30\% at $k{=}50$. In contrast, softmax voting remains stable across the full range of neighbourhood sizes, with accuracy staying around 80.5\% and a standard deviation of only 0.09 percentage points across $k$. At $k{=}50$, softmax voting improves accuracy by 23.09 percentage points over uniform voting.~\autoref{fig:main_kcurve}--a shows this trend, and ~\autoref{tab:knn_stability} summarizes the corresponding stability statistics. Results across all model configurations are reported in Supplementary Materials, Section~I. For BIOSCAN-5M, we further observe that for our model and for nearly all evaluated baselines (\autoref{tab:bioscan_external}), increasing $k$ does not improve accuracy beyond its $k{=}1$ value; softmax voting instead stabilizes accuracy across $k$ rather than surpassing the $k{=}1$ result, with DNABERT-S the only exception showing a modest improvement at larger $k$.

\begin{table}[htp]
\centering
\small
\caption{Stability of the best BIOSCAN-5M configuration (encoder-decoder, +CLS+CE, CLS
representation) across $k \in \{1,3,5,7,10,15,20,25,50\}$, under uniform and softmax
voting. Mean is the average accuracy across all nine values of $k$ tested.}
\label{tab:knn_stability}
\setlength{\tabcolsep}{5pt}
\begin{tabular}{lcccc}
\toprule
Voting & Peak acc.\ (\%) & Mean (\%) & Std.\ (pp) & Range (pp) \\
\midrule
Uniform & 80.65 ($k{=}\phantom{0}1$) & 72.55 & 7.00 & 23.35 \\
Softmax & 80.65 ($k{=}\phantom{0}1$) & 80.53 & 0.09 & \phantom{0}0.27 \\
\bottomrule
\end{tabular}
\end{table}

For UNITE+INSD, the best-performing configuration (encoder-decoder, +CLS+Binary, CLS representation) is more stable across neighbourhood sizes under softmax voting on both test sets. On Yeast, uniform voting drops from a peak accuracy of 74.90\% to 64.45\% at $k{=}50$, whereas softmax voting maintains 76.05\% accuracy at $k{=}50$. On Filamentous, the difference is smaller but follows the same pattern, with softmax voting reaching 65.21\% at $k{=}50$ compared with 60.04\% under uniform voting. \autoref{fig:main_kcurve}--b, and \autoref{fig:main_kcurve}--c show this trend, and \autoref{tab:its_stability} summarizes the corresponding stability statistics. Unlike BIOSCAN-5M, where softmax voting mainly stabilizes rather than improves accuracy beyond $k{=}1$, several baselines on UNITE+INSD (\autoref{tab:its_external}) show gains from increasing $k$ under softmax voting. MycoAI-BERT improves from 70.48\% to 72.81\% on Yeast and from
64.16\% to 66.01\% on Filamentous, while MycoAI-CNN improves from 69.01\% to 71.67\% on Yeast. Our own model shows the same pattern, improving from 73.19\% to 76.43\% on Yeast and from 63.07\% to 65.21\% on Filamentous. Results across all model configurations are reported in Supplementary Materials, Section~I.
\newline
\begin{table}[t]
\centering
\footnotesize
\caption{Stability of the best UNITE+INSD configuration
(encoder-decoder, +CLS+Binary, CLS representation) on the deduplicated
Yeast and Filamentous query pools across
$k \in \{1,3,5,7,10,15,20,25,50\}$ under uniform and softmax voting.
Mean is the average accuracy across all nine values of $k$ tested.}
\label{tab:its_stability}
\setlength{\tabcolsep}{2pt}
\begin{tabular}{llcccc}
\toprule
Test set & Voting & Peak acc.\ (\%) & Mean (\%) & Std.\ (pp) & Range (pp) \\
\midrule
\multirow{2}{*}{Yeast}
  & Uniform & 74.90 ($k{=}7$)  & 72.05 & 3.10 & 10.45 \\
  & Softmax & 76.43 ($k{=}25$) & 75.14 & 1.12 & \phantom{0}3.24 \\
\midrule
\multirow{2}{*}{Filamentous}
  & Uniform & 63.36 ($k{=}7$)  & 62.50 & 0.98 & \phantom{0}3.32 \\
  & Softmax & 65.21 ($k{=}50$) & 64.07 & 0.71 & \phantom{0}2.17 \\
\bottomrule
\end{tabular}
\end{table}

\section{Discussion}
\label{sec:discussion}

This study examines three design choices for DNA barcode foundation models: the pretraining architecture, the use of a global \texttt{[CLS]} representation, and the auxiliary objective used to train that representation. Across arthropod COI and fungal ITS barcodes, the encoder-decoder MAE-LM architecture consistently outperforms the matched encoder-only model in nearly all \(k=1\) comparisons. A well-trained global representation provides further gains. Together, these results show that both the pretraining design and the aggregation of sequence-level information affect the quality of frozen DNA representations.

\textbf{The encoder–decoder model achieves the best-performing configuration for each barcode region and outperforms its matched encoder-only counterpart.}  However, the size of the advantage varies, and the two regions differ in where it is largest. The MAE-LM design may help in two ways that our experiments do not separate:
the encoder never allocates capacity to \texttt{[MASK]} embeddings, and the
decoder interposes between the encoder and the reconstruction loss, so the
encoder representation is not directly coupled to token reconstruction~\citep{dervishi2026crossbert}. On BIOSCAN-5M, the gain is largest for token-based representations without a \texttt{[CLS]} token and narrows once an auxiliary objective directly trains the \texttt{[CLS]} representation. On UNITE+INSD, the pattern is reversed: the gain is small for token-based representations and larger for the \texttt{[CLS]} representation under auxiliary supervision. We do not have an account of why the two regions differ in this way.

\textbf{A global representation is most effective when trained with an appropriate auxiliary objective.} On BIOSCAN-5M, the \texttt{[CLS]} representation performs poorly without auxiliary supervision but improves substantially when taxonomic information is included during pretraining. However, no single objective performs best on both datasets. Direct genus classification gives the strongest result on BIOSCAN-5M, whereas pairwise same-genus prediction performs best on both UNITE+INSD test sets. One possible explanation is that the preferred objective depends on the relationship between the pretraining and evaluation distributions. Direct classification organizes the representation around the genera observed during training and may therefore work well when the downstream data have a similar taxonomic distribution. Pairwise prediction instead learns a relative measure of taxonomic similarity without tying the representation to a fixed set of genus classes. This property may support better transfer when the downstream taxa or sequence distribution differ from those seen during pretraining. More broadly, the  baseline results highlight an implicit value of our methodology. The \texttt{[CLS]} representation benefits not only from auxiliary supervision, but also from pretraining on data that are relevant to the downstream sequence distribution. Across both datasets, models pretrained on the barcode region of interest outperform both general large-scale genomic foundation models and models pretrained on other genomic regions, despite differences in model and pretraining dataset scales.

\textbf{Increasing the capacity of the global token alone does not improve representation quality.} The wider Jumbo token does not outperform the best standard \texttt{[CLS]} configuration, despite its greater capacity. Likewise, changing the taxonomic rank used for auxiliary training does not improve on the best genus-level configuration. Within our evaluation tasks, these results suggest that the training signal applied to the global representation matters more than the size of the \texttt{[CLS]} token or the choice of auxiliary labels.

\textbf{The choice of KNN voting method affects the stability of downstream evaluation.} With uniform voting, accuracy can decrease considerably as the neighbourhood size grows, even for the strongest model configurations. Softmax voting is less sensitive to this choice and maintains strong performance across a wider range of $k$ on both BIOSCAN-5M and UNITE+INSD. This stability is useful for evaluating frozen embeddings, because conclusions about representation quality should not depend heavily on a single choice of neighbourhood size.

\textbf{Summary.} Taken together, the results support the use of encoder-decoder MAE-LM pretraining with an explicitly trained global representation. The auxiliary objective should be chosen for the expected deployment setting: direct taxonomic classification may be preferable when the training and downstream distributions are similar, whereas relational objectives may transfer better when they differ. For frozen-embedding specimen identification, similarity-weighted softmax KNN provides a more stable alternative to uniform voting.
\subsection{Limitations and future work}
\label{sec:limitations}
This study covers only two major barcode regions. Evaluation on additional markers and taxonomic groups, particularly plant barcodes, is needed to determine how broadly these findings generalize. Such evaluation would also show whether the opposing patterns we observe on COI and ITS follow from the markers themselves or from differences in size, label density, and taxonomic breadth between the two datasets. In addition, the UNITE+INSD evaluations use deduplicated query sets obtained by removing samples that overlap with the training or reference set by sample ID or exact barcode sequence. This gives a more rigorous estimate of generalization, but it also changes the taxonomic composition of the original test sets. Larger and more diverse deduplicated ITS benchmarks would support a more complete evaluation. A further limitation concerns the selection of the auxiliary-loss weight (\(\alpha\)). Because neither benchmark provides a validation split matching the unseen-species evaluation setting, we selected \(\alpha\) from a small predefined grid using evaluation-set performance.

Finally, we evaluate frozen representations using KNN classification and zero-shot clustering. This setting measures representation quality without allowing downstream fine-tuning to compensate for weaknesses in pretraining, but it does not show whether the same trends persist after task-specific adaptation. Future work could evaluate BarcodeMAE+ under full supervised fine-tuning and parameter-efficient adaptation, as well as on tasks beyond taxonomic identification and clustering. Auxiliary objectives that use the full taxonomic hierarchy, including hierarchical and multi-rank objectives, also provide a promising direction for future study.
\section{Conclusions}
\label{sec:conclusions}
BarcodeMAE+ provides a practical framework for learning and evaluating frozen DNA barcode representations. Our results show that representation quality depends not only on the pretraining architecture, but also on how the global representation is trained and how the resulting embeddings are evaluated. The consistent performance of encoder-decoder MAE-LM across COI and ITS supports its use as a foundation for future barcode models, while the variation among auxiliary objectives shows that their selection should reflect the target domain. Extending this framework to additional barcode regions, taxonomic groups, and downstream adaptation settings will help establish how broadly these design principles apply.

\FloatBarrier
\section{Code Availability}
All code used in this study is publicly available on GitHub at \href{https://github.com/bioscan-ml/BarcodeMAE-plus}{https://github.com/bioscan-ml/BarcodeMAE-plus}

\section{Data Availability}
All data used in this study is publicly available. BIOSCAN-5M metadata and DNA barcode sequences are available from the BIOSCAN-5M GitHub repository \href{https://github.com/bioscan-ml/BIOSCAN-5M}{https://github.com/bioscan-ml/BIOSCAN-5M}. The UNITE+INSD fungal ITS dataset is available from the MycoAI GitHub repository \href{https://github.com/MycoAI/MycoAI}{https://github.com/MycoAI/MycoAI}.

\section{Competing interests}
No competing interest is declared.

\section{Funding}
We acknowledge the support of the Government of Canada's New Frontiers in Research Fund \href{https://sshrc-crsh.canada.ca/funding-financement/nfrf-fnfr/transformation/2020/award_recipients-titulaires_subvention-eng.aspx}{Award No. NFRFT-2020-00073} and the \href{http://abcresearchcenter.org/}{AI and Biodiversity Change (ABC) Global Center}, which is funded by the US National Science Foundation under \href{https://www.nsf.gov/awardsearch/showAward?AWD_ID=2330423&HistoricalAwards=false}{Award No. 2330423} and Natural Sciences and Engineering Research Council of Canada under \href{https://www.nserc-crsng.gc.ca/ase-oro/Details-Detailles_eng.asp?id=782440}{Award No. 585136}. This research was supported, in part, by the Province of Ontario and the Government of Canada through the Canadian Institute for Advanced Research (CIFAR), and \href{https://vectorinstitute.ai/partnerships/current-partners/}{companies sponsoring} the Vector Institute.
GWT and AXC are supported by the Natural Sciences and Engineering Research Council of Canada (NSERC) and the Canada CIFAR AI Chairs program. GWT is also supported by the Canada Research Chairs program (grant~CRC-2021-00561). GWT and SCL are supported by the INSPIRE (Integrated Network for the Surveillance of Pathogens: Increasing REsilience and capacity in Canada's pandemic response) project funded through the Canada Biomedical Research Fund (grant CBRF2-2023-00008), the Biomedical Research Infrastructure Fund, and the Ontario Research Fund. LK is also supported by NSERC Discovery Grant RGPIN-2023-03663.

\section{Authors' Contributions}
Monireh Safari preprocessed and deduplicated the datasets, preparing them for evaluation. Monireh Safari and Pablo Millan Arias designed and implemented BarcodeMAE+. Monireh Safari conducted the BarcodeMAE+ experiments and the DNA baseline experiments. Monireh Safari and Pablo Millan Arias wrote the manuscript and prepared the figures. Scott C. Lowe, Graham W. Taylor, Angel X. Chang, and Lila Kari provided guidance on the experimental design and edited the manuscript. All authors reviewed and approved the final manuscript.

\bibliographystyle{plainnat}
\bibliography{Sections/references}

\appendix
\makeatletter
\renewcommand{\@seccntformat}[1]{}
\makeatother
\section*{Appendices}

\section{A. Dataset Taxonomic Distribution}
\label{app:taxdist}

\begin{figure*}[h]
\centering
\begin{subfigure}[t]{0.48\textwidth}
\centering
\includegraphics[width=\linewidth]{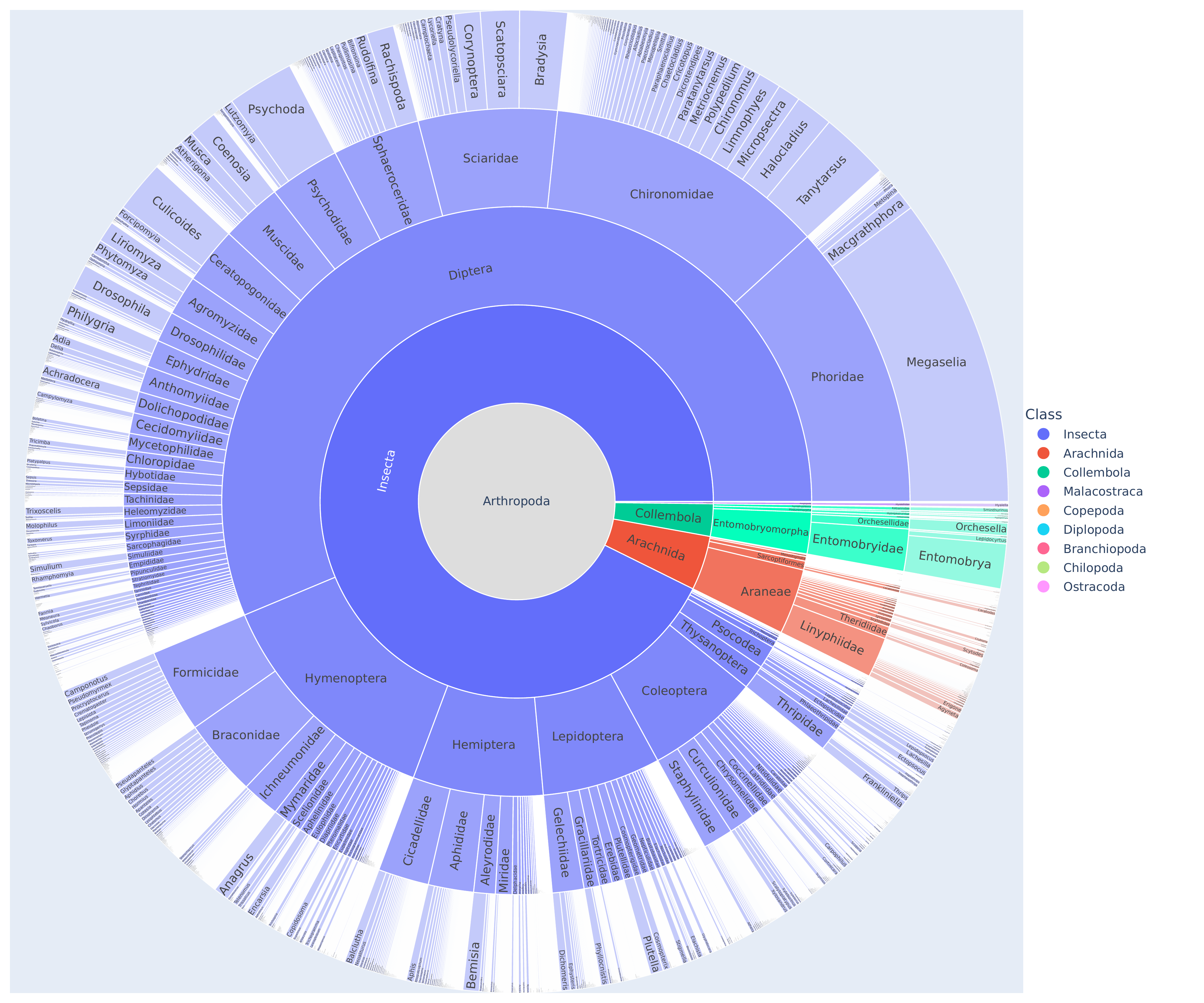}
\caption{KNN reference set (Seen/train, 117,991 specimens)}
\end{subfigure}
\hfill
\begin{subfigure}[t]{0.48\textwidth}
\centering
\includegraphics[width=\linewidth]{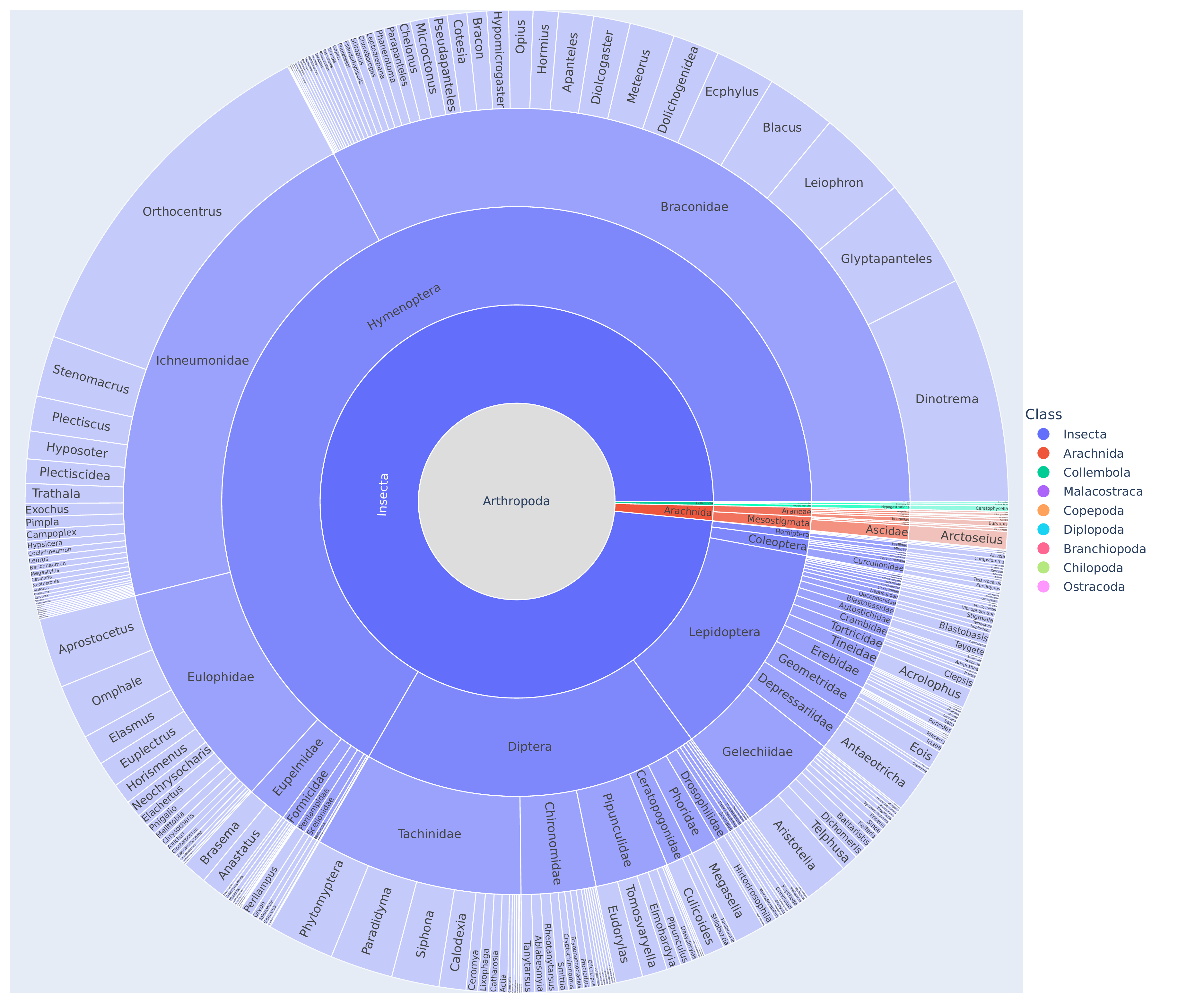}
\caption{Query set (Unseen, 3,396 specimens)}
\end{subfigure}
\caption{BIOSCAN-5M hierarchical taxonomic distribution. Each ring is one
taxonomic rank, from phylum (innermost) through class, order, family, to
genus (outermost); wedge angular width is proportional to the number of
specimens at that taxon. Since BIOSCAN-5M covers only arthropods, the
innermost ring is a single wedge (phylum Arthropoda); colour and the legend
denote class, the coarsest rank with more than one value in this dataset,
with descendant ranks shaded progressively lighter shades of their class's
colour moving outward.}
\label{fig:appendix_taxdist_bioscan5m}
\end{figure*}

\begin{figure*}[h]
\centering
\begin{subfigure}[t]{0.32\textwidth}
\centering
\includegraphics[width=\linewidth]{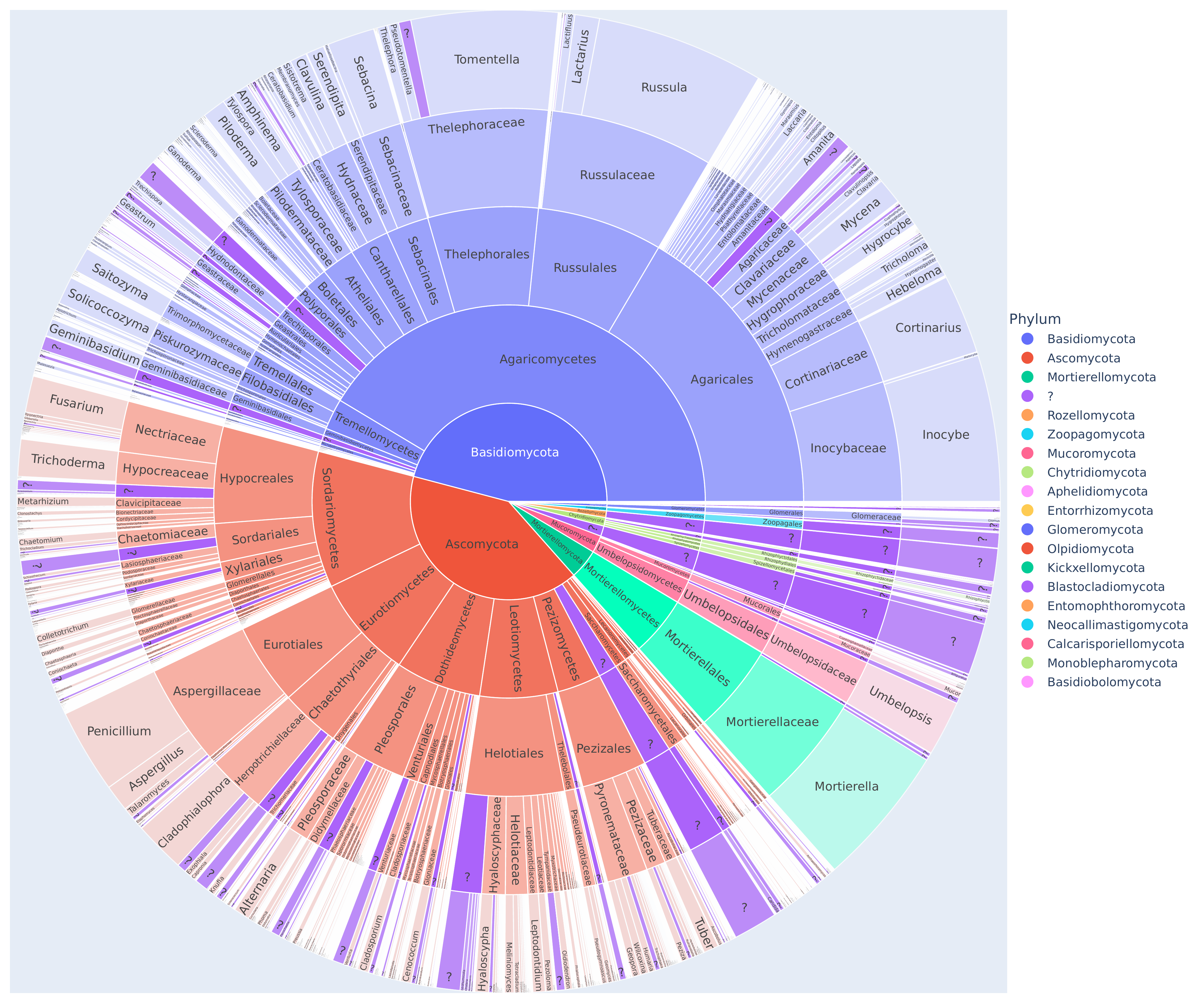}
\caption{KNN reference set (train, 5.23M sequences)}
\end{subfigure}
\hfill
\begin{subfigure}[t]{0.32\textwidth}
\centering
\includegraphics[width=\linewidth]{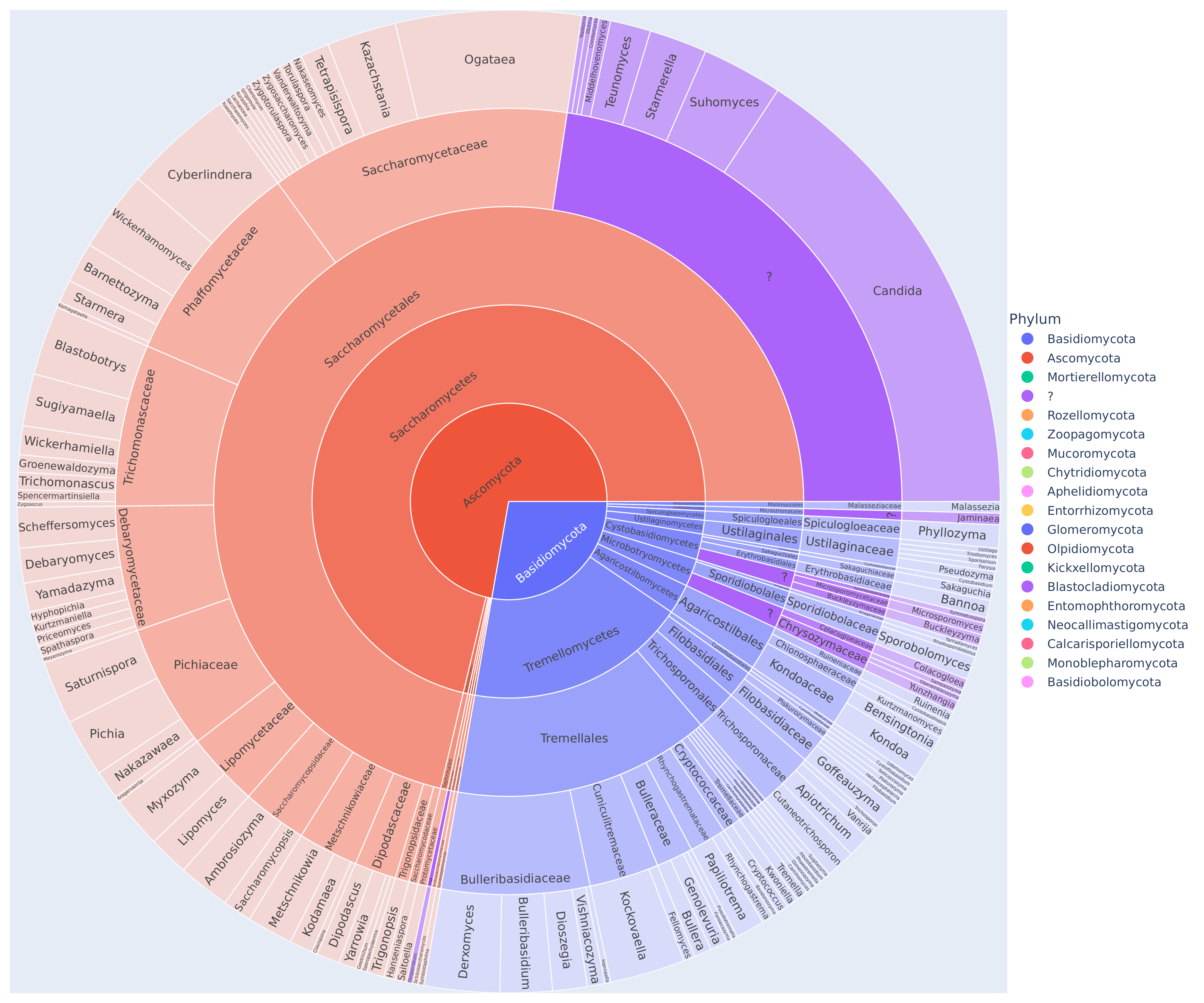}
\caption{Yeast query set (deduplicated, 526 specimens)}
\end{subfigure}
\hfill
\begin{subfigure}[t]{0.32\textwidth}
\centering
\includegraphics[width=\linewidth]{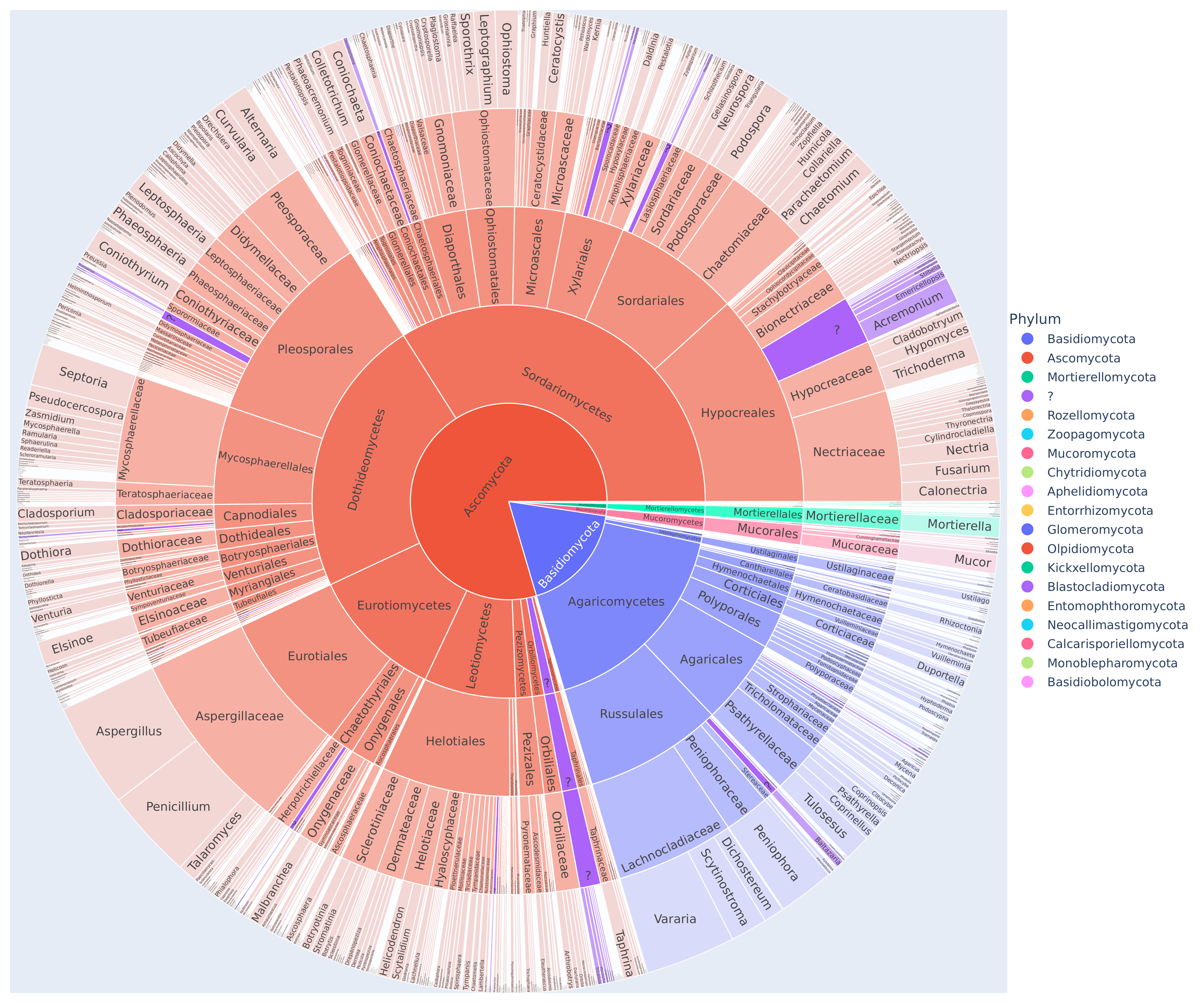}
\caption{Filamentous query set (deduplicated, 3,136 specimens)}
\end{subfigure}
\caption{UNITE+INSD hierarchical taxonomic distribution. As in
\autoref{fig:appendix_taxdist_bioscan5m}, each ring is one taxonomic
rank, from phylum (innermost) through class, order, family, to genus
(outermost), with wedge angular width proportional to the number of
specimens at that taxon. Colour and the legend denote phylum here (rather
than class, as for BIOSCAN-5M), since UNITE+INSD spans multiple fungal
phyla rather than a single one; descendant ranks are shaded progressively
lighter shades of their phylum's colour moving outward. The two query sets
are the deduplicated, genus-level query populations actually used for
evaluation (\autoref{tab:appendix_its_leakage}), not the raw released
test sets.}
\label{fig:appendix_taxdist_its5m}
\end{figure*}

\autoref{fig:appendix_taxdist_bioscan5m} shows the hierarchical taxonomic
distribution of BIOSCAN-5M's KNN reference set and query set, and
~\autoref{fig:appendix_taxdist_its5m} shows the same for UNITE+INSD.

~\autoref{tab:appendix_its_leakage} summarizes the statistics of UNITE+INSD original training and test sets, and the deduplicated evaluation sets described
in the main text. We report the number of sequences, unique species, and unique genera in the
training set, the original released test set, and the deduplicated genus-level query set
actually used for evaluation (526 Yeast / 3,136 Filamentous, as reported in the
Experimental Setup).

\begin{table*}[h]
\centering
\small
\caption{UNITE+INSD for the Yeast and Filamentous test sets. Training set
is the 5.23M-sequence pretraining corpus. Our test set is the deduplicated genus-level
query set used for evaluation: exact-sequence duplicates of the training set are
removed, the remaining specimens are restricted to those with a resolved species label
that does not appear in the training set at all (unseen species).}
\label{tab:appendix_its_leakage}
\setlength{\tabcolsep}{4pt}
\begin{tabular}{llccc}
\toprule
Test set & Metric & Training set & Original test set & Our test set \\
\midrule
\multirow{3}{*}{Yeast (Test 1)}
  & Sequences       & 5,230,185 & 4,247 & 526 \\
  & Unique species  & 14,742    & 1,126 & 476 \\
  & Unique genera   & 3,695     & 203   & 122 \\
\midrule
\multirow{3}{*}{Filamentous (Test 2)}
  & Sequences       & 5,230,185 & 10,873 & 3,136 \\
  & Unique species  & 14,742    & 5,415  & 2,348 \\
  & Unique genera   & 3,695     & 1,642  & 827 \\
\bottomrule
\end{tabular}
\end{table*}

\FloatBarrier
\section{B. BIOSCAN-5M and UNITE+INSD: Family and BIN Auxiliary Label Results}
\label{app:taxlevel}

This section repeats the main experiment, changing only which labels supervise the auxiliary-task: \textit{family} and \textit{BIN} for BIOSCAN-5M (\autoref{tab:appendix_bioscan_family} and \autoref{tab:appendix_bioscan_bin}, respectively), and \textit{family} for UNITE+INSD (\autoref{tab:appendix_its_family}), which does not provide BIN labels. For BIOSCAN-5M, 92.9\% of pretraining sequences carry a family label and 99.5\% carry a BIN label, compared with 15.9\% at genus rank; for UNITE+INSD, 84.2\% carry a family label, compared with 80.3\% at genus rank. The evaluation task remains unchanged: genus-level 1-NN accuracy for both datasets and zero-shot BIN-reconstruction AMI for BIOSCAN-5M, as in the main results. Bold indicates the best value in each column, and underlining indicates the second-best.

Encoder-decoder remains the best overall configuration. However, none of these configurations outperform the genus-level auxiliary labels' 1-NN accuracy used in the main text (80.65\% for BIOSCAN-5M, 73.19\% for UNITE+INSD Yeast, and 63.07\% for UNITE+INSD Filamentous), indicating that genus is the most effective rank for the auxiliary labels. For 1-NN evaluation, CE remains the best auxiliary objective at the family rank, but with BIN labels it is surpassed by both Binary and Triplet (65.78\% versus 72.56\% and 72.61\%, respectively). Thus, CE's advantage does not extend to the finest rank. For UNITE+INSD at family level, CE and Binary both perform well and are close to each other (71.67\% versus 71.29\% Yeast, and 58.99\% versus 60.52\% Filamentous), with Triplet noticeably behind.

\begin{table}[h]
\centering
\small
\caption{BIOSCAN-5M results with family-level auxiliary labels. Genus-level 1-NN
accuracy (\%) and zero-shot BIN reconstruction AMI (\%), for Binary,
Triplet, and CE. \textbf{Bold} marks the best 
value in each column; \underline{underline} marks the second-best.}
\label{tab:appendix_bioscan_family}
\setlength{\tabcolsep}{5pt}
\begin{tabular}{llcccc}
\toprule
 & & \multicolumn{2}{c}{Encoder-decoder} & \multicolumn{2}{c}{Encoder-only} \\
\cmidrule(lr){3-4} \cmidrule(lr){5-6}
Aux & Repr & 1-NN Acc.\ (\%) & ZSC AMI (\%) & 1-NN Acc.\ (\%) & ZSC AMI (\%) \\
\midrule
Binary  & Tokens  & 65.14 & \underline{74.67} & 48.85 & 55.28 \\
Binary  & CLS     & \underline{71.23} & \textbf{74.86} & 55.77 & 71.39 \\
Binary  & Tokens+CLS & 65.37 & 74.35 & 48.91 & 55.52 \\
Triplet (hard) & Tokens  & 64.08 & 74.30 & 50.53 & 55.02 \\
Triplet (hard) & CLS     & 57.77 & 73.33 & 58.57 & 72.59 \\
Triplet (hard) & Tokens+CLS & 64.10 & 74.13 & 50.65 & 55.13 \\
CE      & Tokens  & 67.31 & 74.55 & 52.09 & 57.29 \\
CE      & CLS     & \textbf{71.26} & 74.30 & 60.87 & 72.89 \\
CE      & Tokens+CLS & 67.26 & 74.27 & 52.56 & 57.23 \\
\bottomrule
\end{tabular}
\end{table}

\begin{table}[h]
\centering
\small
\caption{BIOSCAN-5M results with BIN auxiliary labels. Same evaluation task as
\autoref{tab:appendix_bioscan_family}. \textbf{Bold}
marks the best value in each column; \underline{underline} marks the second-best.}
\label{tab:appendix_bioscan_bin}
\setlength{\tabcolsep}{5pt}
\begin{tabular}{llcccc}
\toprule
 & & \multicolumn{2}{c}{Encoder-decoder} & \multicolumn{2}{c}{Encoder-only} \\
\cmidrule(lr){3-4} \cmidrule(lr){5-6}
Aux & Repr & 1-NN Acc.\ (\%) & ZSC AMI (\%) & 1-NN Acc.\ (\%) & ZSC AMI (\%) \\
\midrule
Binary  & Tokens  & 68.90 & 74.93 & 60.45 & 57.46 \\
Binary  & CLS     & \underline{72.56} & \underline{75.27} & 62.93 & 73.38 \\
Binary  & Tokens+CLS & 68.99 & \textbf{75.29} & 60.51 & 57.64 \\
Triplet (hard) & Tokens  & \underline{72.56} & 74.82 & 63.43 & 56.50 \\
Triplet (hard) & CLS     & 68.82 & 74.39 & 71.82 & 74.01 \\
Triplet (hard) & Tokens+CLS & \textbf{72.61} & 74.71 & 63.49 & 56.41 \\
CE      & Tokens  & 65.69 & 74.12 & 46.91 & 57.97 \\
CE      & CLS     & 59.84 & 73.40 & 57.33 & 73.51 \\
CE      & Tokens+CLS & 65.78 & 74.53 & 46.91 & 57.86 \\
\bottomrule
\end{tabular}
\end{table}

\begin{table}[h]
\centering
\small
\caption{UNITE+INSD results with family-level auxiliary labels. Genus-level 1-NN
accuracy (\%) on the deduplicated Yeast and Filamentous query pools, for auxiliary
pairwise/triplet/CE labels drawn from family rank. \textbf{Bold} marks the best value in
each column; \underline{underline} marks the second-best.}
\label{tab:appendix_its_family}
\setlength{\tabcolsep}{5pt}
\begin{tabular}{llcccc}
\toprule
 & & \multicolumn{2}{c}{Encoder-decoder} & \multicolumn{2}{c}{Encoder-only} \\
\cmidrule(lr){3-4} \cmidrule(lr){5-6}
Aux & Repr & Yeast 1-NN (\%) & Filamentous 1-NN (\%) & Yeast 1-NN (\%) & Filamentous 1-NN (\%) \\
\midrule
Binary  & Tokens  & 59.89 & 52.17 & 56.27 & 52.10 \\
Binary  & CLS     & \underline{71.29} & \textbf{60.52} & 63.69 & 55.96 \\
Binary  & Tokens+CLS & 60.27 & 52.58 & 57.03 & 52.20 \\
Triplet (hard) & Tokens  & 58.37 & 51.66 & 55.51 & 50.96 \\
Triplet (hard) & CLS     & 67.11 & 57.21 & 64.64 & 55.13 \\
Triplet (hard) & Tokens+CLS & 58.56 & 51.79 & 55.70 & 51.05 \\
CE      & Tokens  & 61.22 & 52.55 & 56.20 & 50.90 \\
CE      & CLS     & \textbf{71.67} & \underline{58.99} & 62.10 & 51.20 \\
CE      & Tokens+CLS & 61.79 & 52.68 & 58.40 & 52.84 \\
\bottomrule
\end{tabular}
\end{table}

\FloatBarrier
\section{C. Softmax Temperature Experiment}
\label{app:temperature_ablation}

\begin{figure}[h]
\centering
\includegraphics[width=\linewidth]{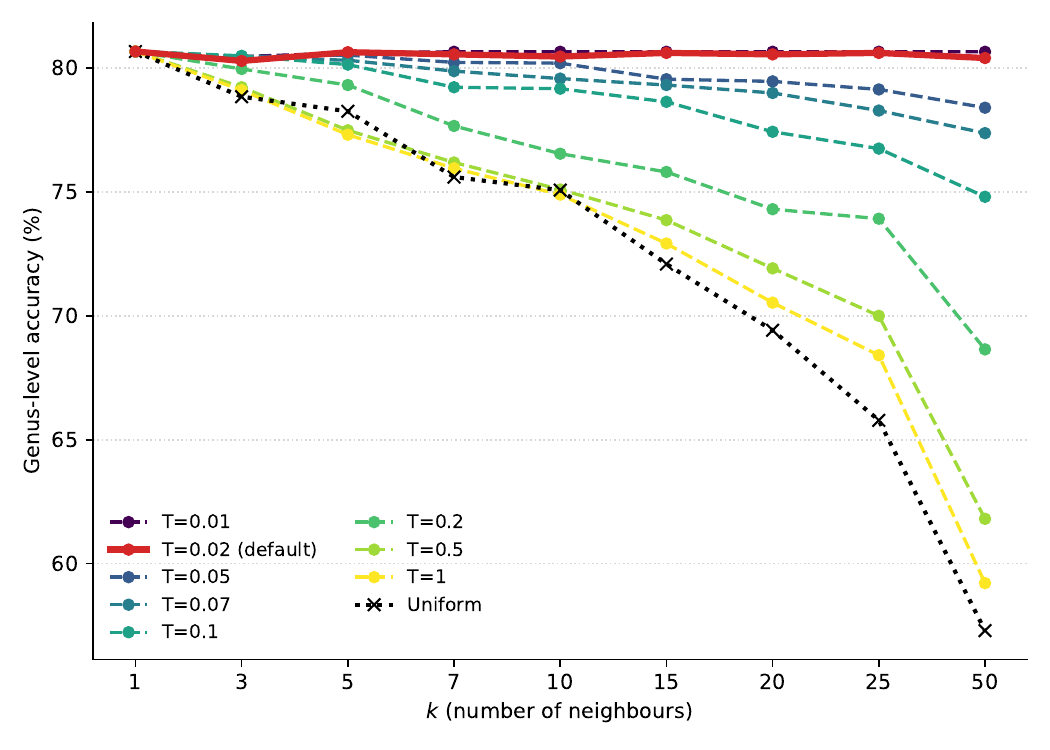}
\caption{BIOSCAN-5M genus-level KNN accuracy vs.\ $k$ under softmax voting, for the best
configuration (encoder-decoder, +CLS+CE, CLS representation), at each tested
temperature, plus uniform voting (black dotted) for reference.
$T=0.02$ is highlighted in red; lower
temperatures are closer to winner-take-all voting, higher
temperatures (warmer colours) are closer to uniform voting, and the $T=1.0$ curve is
visibly converging toward the uniform-voting curve as $k$ grows.}
\label{fig:appendix_temperature_ablation}
\end{figure}

\begin{figure}[h]
\centering
\includegraphics[width=\linewidth]{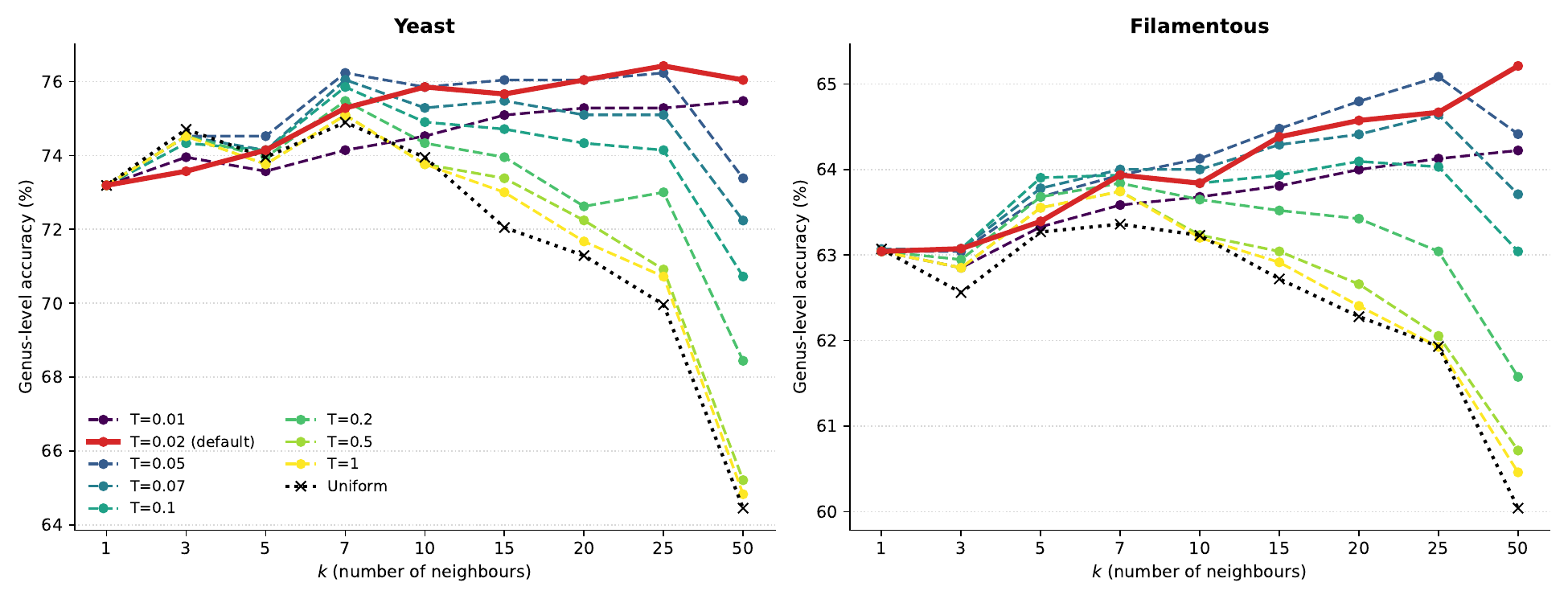}
\caption{UNITE+INSD genus-level KNN accuracy vs.\ $k$ under softmax voting, for the best
configuration (encoder-decoder, +CLS+Binary, CLS representation), at each tested
temperature, plus uniform voting (black dotted) for reference, on
the deduplicated Yeast (left) and Filamentous (right) query pools. $T=0.02$ (the value
adopted throughout the main text) is highlighted in red.}
\label{fig:appendix_temperature_ablation_its}
\end{figure}

We test $T \in \{0.01, 0.02, 0.05, 0.07, 0.1, 0.2, 0.5, 1.0\}$ for the best BIOSCAN-5M configuration (encoder-decoder, +CLS+CE, CLS representation) and the best UNITE+INSD configuration (encoder-decoder, +CLS+Binary, CLS representation) on the deduplicated Yeast and Filamentous query pools (\autoref{fig:appendix_temperature_ablation}, \autoref{fig:appendix_temperature_ablation_its}, \autoref{tab:appendix_temperature_ablation}, \autoref{tab:appendix_temperature_ablation_its}). $T = 0.02$ gives the best peak accuracy on every test set and the best mean on Yeast; on Filamentous, $T=0.05$ attains a marginally higher mean (64.07\% vs.\ 64.01\%), and on BIOSCAN-5M, $T=0.01$ attains a marginally higher mean (80.61\% vs.\ 80.53\%). We adopt $T = 0.02$ as the default throughout the main text for its consistently best peak accuracy and near-identical mean performance.

\begin{table}[h]
\centering
\small
\caption{BIOSCAN-5M softmax-voting stability statistics for each tested temperature,
across $k \in \{1,3,5,7,10,15,20,25,50\}$. Mean is the average accuracy across all nine
values of $k$ tested. $T=0.02$ is
the value adopted throughout the main text; $T=0.07$ is DINOv2's original default,
included for comparison. \textbf{Bold} marks the best (highest for Peak/Mean, lowest for
Std./Range) value in each column; \underline{underline} marks the second-best.}
\label{tab:appendix_temperature_ablation}
\setlength{\tabcolsep}{5pt}
\begin{tabular}{lcccc}
\toprule
$T$ & Peak acc.\ (\%) & Mean (\%) & Std.\ (pp) & Range (pp) \\
\midrule
0.01                  & \underline{80.64} ($k{=}\phantom{0}1$) & \textbf{80.61} & \textbf{0.06} & \textbf{0.19} \\
0.02 (default)        & \textbf{80.65} ($k{=}\phantom{0}1$) & \underline{80.53} & \underline{0.09} & \underline{0.27} \\
0.05                  & \textbf{80.65} ($k{=}\phantom{0}1$) & 79.83 & 0.72 & 2.27 \\
0.07 (DINOv2 default) & \textbf{80.65} ($k{=}\phantom{0}1$) & 79.41 & 1.02 & 3.29 \\
0.1                   & \textbf{80.65} ($k{=}\phantom{0}1$) & 78.58 & 1.83 & 5.86 \\
0.2                   & \textbf{80.65} ($k{=}\phantom{0}1$) & 76.30 & 3.52 & 12.01 \\
0.5                   & \textbf{80.65} ($k{=}\phantom{0}1$) & 74.02 & 5.36 & 18.85 \\
1.0                   & \textbf{80.65} ($k{=}\phantom{0}1$) & 73.21 & 6.18 & 21.44 \\
\bottomrule
\end{tabular}
\end{table}

\begin{table}[h]
\centering
\footnotesize
\caption{UNITE+INSD softmax-voting stability statistics for each tested temperature, for
the best configuration (encoder-decoder, +CLS+Binary, CLS representation) on the
deduplicated Yeast and Filamentous query pools, across $k \in \{1,3,5,7,10,15,20,25,50\}$.
Mean is the average accuracy across all nine values of $k$ tested. $T=0.02$ (matches
the corresponding main-text table's softmax rows) is the value adopted throughout the main
text; $T=0.07$ is DINOv2's original default, included for comparison. \textbf{Bold} marks the best
(highest for Peak/Mean, lowest for Std./Range) value in each column, per test set;
\underline{underline} marks the second-best.}
\label{tab:appendix_temperature_ablation_its}
\setlength{\tabcolsep}{4pt}
\begin{tabular}{llcccc}
\toprule
Test set & $T$ & Peak acc.\ (\%) & Mean (\%) & Std.\ (pp) & Range (pp) \\
\midrule
\multirow{8}{*}{Yeast (Test 1)}
  & 0.01                  & 75.48 ($k{=}50$) & 74.50 & \textbf{0.78} & \textbf{2.28} \\
  & 0.02 (default)        & \textbf{76.43} ($k{=}25$) & \textbf{75.14} & \underline{1.12} & 3.24 \\
  & 0.05                  & \underline{76.24} ($k{=}7$)  & \underline{75.12} & 1.16 & \underline{3.04} \\
  & 0.07 (DINOv2 default) & 76.05 ($k{=}7$)  & 74.57 & 1.14 & 3.81 \\
  & 0.1                   & 75.86 ($k{=}7$)  & 74.04 & 1.35 & 5.13 \\
  & 0.2                   & 75.48 ($k{=}7$)  & 73.28 & 1.90 & 7.03 \\
  & 0.5                   & 75.10 ($k{=}7$)  & 72.45 & 2.81 & 9.89 \\
  & 1.0                   & 75.10 ($k{=}7$)  & 72.29 & 2.93 & 10.27 \\
\midrule
\multirow{8}{*}{Filamentous (Test 2)}
  & 0.01                  & 64.22 ($k{=}50$) & 63.63 & \underline{0.45} & \underline{1.37} \\
  & 0.02 (default)        & \textbf{65.21} ($k{=}50$) & \underline{64.01} & 0.71 & 2.17 \\
  & 0.05                  & \underline{65.08} ($k{=}25$) & \textbf{64.07} & 0.68 & 2.04 \\
  & 0.07 (DINOv2 default) & 64.64 ($k{=}25$) & 63.89 & 0.52 & 1.57 \\
  & 0.1                   & 64.09 ($k{=}20$) & 63.66 & \textbf{0.43} & \textbf{1.05} \\
  & 0.2                   & 63.84 ($k{=}7$)  & 63.19 & 0.65 & 2.26 \\
  & 0.5                   & 63.74 ($k{=}7$)  & 62.77 & 0.86 & 3.03 \\
  & 1.0                   & 63.74 ($k{=}7$)  & 62.68 & 0.94 & 3.28 \\
\bottomrule
\end{tabular}
\end{table}

\FloatBarrier
\section{D. Best Temperature for External Baselines}
\label{app:baseline_temperature}

As described in the Experimental Setup, the Softmax KNN results reports each baseline's
best accuracy across the full $k \times T$ grid ($k \in \{1,3,5,7,10,15,20,25,50\}$,
$T \in \{0.01,0.02,0.05,0.07,0.1,0.2,0.5,1.0\}$); a dash in those tables means $k{=}1$
already performs best, at which point temperature has no effect. \autoref{tab:appendix_baseline_temperature}
reports the specific $(T,k)$ pair behind each non-dash entry.

\begin{table}[h]
\centering
\small
\caption{Best temperature and $k$ for external baselines where softmax
voting at $k>1$ improves over $k{=}1$.}
\label{tab:appendix_baseline_temperature}
\setlength{\tabcolsep}{5pt}
\begin{tabular}{lllccc}
\toprule
Dataset & Test set & Model & Best $T$ & Best $k$ & Accuracy (\%) \\
\midrule
BIOSCAN-5M & --- & DNABERT-S~\cite{zhou2024dnabertS} & 0.01 & 20 & 20.58 \\
\midrule
UNITE+INSD & Yeast       & BarcodeMamba+ (large)~\cite{gao2025barcodemambaplus} & 0.01 & 7  & 62.54 \\
UNITE+INSD & Filamentous & BarcodeMamba+~\cite{gao2025barcodemambaplus}         & 0.01 & 5  & 57.04 \\
UNITE+INSD & Filamentous & BarcodeMamba+ (large)~\cite{gao2025barcodemambaplus} & 0.01 & 5  & 56.31 \\
UNITE+INSD & Filamentous & DNABERT-2~\cite{zhou2023dnabert2}                    & 0.01 & 3  & 54.15 \\
UNITE+INSD & Yeast       & DNABERT-S~\cite{zhou2024dnabertS}                    & 0.01 & 5  & 50.95 \\
UNITE+INSD & Filamentous & DNABERT-S~\cite{zhou2024dnabertS}                    & 0.01 & 5  & 52.17 \\
UNITE+INSD & Yeast       & MycoAI-BERT~\cite{MycoAI}                            & 0.05 & 5  & 72.81 \\
UNITE+INSD & Filamentous & MycoAI-BERT~\cite{MycoAI}                            & 0.05 & 15 & 66.01 \\
UNITE+INSD & Yeast       & MycoAI-CNN~\cite{MycoAI}                             & 0.1  & 5  & 71.67 \\
UNITE+INSD & Filamentous & MycoAI-CNN~\cite{MycoAI}                             & 0.05 & 5  & 63.93 \\
\bottomrule
\end{tabular}
\end{table}

\FloatBarrier
\section{E. UNITE+INSD: Results on the Original (Non-Deduplicated) Test Sets}
\label{app:its_original}

The corresponding table in the main text reports results on the deduplicated Yeast and
Filamentous query pools (526 and 3,136 specimens respectively, after removing
sample-ID and exact-sequence overlap with the training set; see
\autoref{tab:appendix_its_leakage}). Here we report genus-level 1-NN accuracy for the
same set of baselines on the original, non-deduplicated published test sets (4,247 Yeast
and 10,873 Filamentous specimens), for direct comparison with results reported elsewhere
in the literature on these standard splits, in \autoref{tab:appendix_its_original}.
BarcodeMAE+ achieves the best Yeast accuracy (96.15\%); MycoAI-BERT is best on
Filamentous (83.09\%).

\begin{table*}[h]
\centering
\footnotesize
\caption{Comparison with baselines on \textbf{UNITE+INSD}'s original (non-deduplicated)
published Yeast and Filamentous test sets, evaluated using frozen embeddings. We report
genus-level 1-NN accuracy for each test set. Training task abbreviations are, NTP = next-token
prediction, MLM = masked language modelling, CL = contrastive learning, SC = supervised
classification, MLM+Binary = masked language modelling with a same-genus pairwise auxiliary
objective. \textbf{Bold} marks the best
(highest) value in each column; \underline{underline} marks the second-best.}
\label{tab:appendix_its_original}
\setlength{\tabcolsep}{5pt}
\begin{tabular}{llllcc}
\toprule
Architecture & Training data & Model & Training task & Yeast 1-NN (\%) & Filamentous 1-NN (\%) \\
\midrule

\multirow{4}{*}{State space}
  & Human genome & HyenaDNA-tiny~\cite{nguyen2024hyenadna}        & NTP & 90.87 & 67.47 \\
  & Human genome & Caduceus-PS-1k~\cite{schiff2024caduceus}       & MLM & 89.83 & 64.27 \\
  & UNITE+INSD   & BarcodeMamba+~\cite{gao2025barcodemambaplus}   & NTP & 94.62 & 78.51 \\
  & UNITE+INSD   & BarcodeMamba+ (large)~\cite{gao2025barcodemambaplus} & NTP & 94.46 & 78.87 \\
\midrule

\multirow{6}{*}{Encoder-only}
  & Multi-species DNA & DNABERT-2~\cite{zhou2023dnabert2}              & MLM & 94.43 & 77.90 \\
  & Multi-species DNA & DNABERT-S~\cite{zhou2024dnabertS}              & CL  & 93.03 & 76.94 \\
  & Multi-species DNA & Nucleotide Transformer~\cite{dalla2024nucleotide} & MLM & 91.10 & 68.50 \\
  & Canada-1.5M       & BarcodeBERT~\cite{millan2023barcodebert}       & MLM & 88.61 & 61.02 \\
  & Human genome      & GENA-LM~\cite{fishman2025genalm}               & MLM & 94.52 & 77.82\\
  & UNITE+INSD        & MycoAI-BERT~\cite{MycoAI}                      & SC  & \underline{95.92} & \textbf{83.09} \\
\midrule

CNN
  & UNITE+INSD & MycoAI-CNN~\cite{MycoAI} & SC & 95.58 & \underline{82.16} \\
\midrule

Encoder-decoder
  & UNITE+INSD & \textbf{BarcodeMAE+ (Ours)} & MLM+Binary & \textbf{96.15} & 81.92 \\

\bottomrule
\end{tabular}
\end{table*}

\FloatBarrier
\section{F. Zero-Shot BIN Reconstruction Across All Representations}
\label{app:zsc_all_repr}

\autoref{tab:appendix_zsc_all_repr} reports zero-shot BIN-reconstruction AMI across all configurations and representations, including both Triplet mining strategies. The encoder-decoder model with \texttt{+CLS+CE} achieves the best result using the CLS representation (75.79\%), followed by its Tokens representation (75.15\%). Without auxiliary supervision, CLS performs poorly, whereas Binary, Triplet, and CE supervision substantially improve it. As with genus-level 1-NN accuracy, random-mining Triplet degrades the encoder-decoder CLS representation (69.80\%) below its encoder-only counterpart (72.85\%), the same reversal reported in the main text. Overall, encoder-decoder pretraining produces strong token representations, while auxiliary supervision is essential for learning an effective CLS representation.

\begin{table*}[h]
\centering
\small
\caption{Zero-shot BIN reconstruction AMI (\%) on BIOSCAN-5M for all
representations of every model configuration, including both Triplet mining
strategies. \textbf{Bold} marks the best (highest) value;
\underline{underline} marks the second-best.}
\label{tab:appendix_zsc_all_repr}
\setlength{\tabcolsep}{5pt}
\begin{tabular}{llcc}
\toprule
 & & Encoder-decoder & Encoder-only \\
\cmidrule(lr){3-3} \cmidrule(lr){4-4}
Configuration & Representation & BIN AMI & BIN AMI \\
\midrule
(no CLS)     & Tokens  & 74.32 & 55.65 \\
+CLS         & Tokens  & 74.67 & 54.40 \\
+CLS         & CLS     & 54.94 & 51.78 \\
+CLS         & Tokens+CLS & 74.79 & 54.40 \\
+CLS+Binary  & Tokens  & 74.72 & 55.79 \\
+CLS+Binary  & CLS     & 75.02 & 73.39 \\
+CLS+Binary  & Tokens+CLS & 74.82 & 56.16 \\
+CLS+Triplet (hard)   & Tokens  & 74.50 & 54.65 \\
+CLS+Triplet (hard)   & CLS     & 74.31 & 72.76 \\
+CLS+Triplet (hard)   & Tokens+CLS & 74.50 & 54.99 \\
+CLS+Triplet (random) & Tokens  & 74.75 & 54.96 \\
+CLS+Triplet (random) & CLS     & 69.80 & 72.85 \\
+CLS+Triplet (random) & Tokens+CLS & 74.62 & 54.92 \\
+CLS+CE      & Tokens  & \underline{75.15} & 55.94 \\
+CLS+CE      & CLS     & \textbf{75.79} & 74.45 \\
+CLS+CE      & Tokens+CLS & 75.11 & 55.97 \\
\bottomrule
\end{tabular}
\end{table*}

\FloatBarrier
\section{G. Auxiliary Objectives and Representations: Full Genus-Level 1-NN Results}
\label{app:fig2_full}

\autoref{tab:appendix_fig2_bioscan} and \autoref{tab:appendix_fig2_its} report the genus-level 1-NN accuracy for every architecture, auxiliary objective, and sequence representation, together with the randomly-initialized encoder baseline.

\begin{table*}[h]
\centering
\small
\caption{Genus-level 1-NN accuracy (\%) on BIOSCAN-5M for every configuration and
representation, including randomly-initialized encoder baseline. \textbf{Bold} marks the best
value in each column; \underline{underline} marks the second-best.}
\label{tab:appendix_fig2_bioscan}
\setlength{\tabcolsep}{5pt}
\begin{tabular}{llcc}
\toprule
 & & Encoder-decoder & Encoder-only \\
Configuration & Representation & 1-NN Acc.\ (\%) & 1-NN Acc.\ (\%) \\
\midrule
(no CLS) & Tokens & 66.73 & 51.27 \\
+CLS & Tokens & 68.76 & 55.39 \\
+CLS & CLS & 47.53 & 47.53 \\
+CLS & Tokens+CLS & 68.52 & 55.42 \\
+CLS+Binary & Tokens & 69.05 & 58.19 \\
+CLS+Binary & CLS & 74.76 & \underline{67.26} \\
+CLS+Binary & Tokens+CLS & 69.26 & 58.36 \\
+CLS+Triplet(hard) & Tokens & 70.44 & 57.60 \\
+CLS+Triplet(hard) & CLS & \underline{74.85} & 66.02 \\
+CLS+Triplet(hard) & Tokens+CLS & 70.47 & 57.95 \\
+CLS+Triplet(random) & Tokens & 67.76 & 55.33 \\
+CLS+Triplet(random) & CLS & 50.32 & 62.37 \\
+CLS+Triplet(random) & Tokens+CLS & 67.70 & 55.45 \\
+CLS+CE & Tokens & 73.35 & 60.84 \\
+CLS+CE & CLS & \textbf{80.65} & \textbf{71.67} \\
+CLS+CE & Tokens+CLS & 73.41 & 61.07 \\
\midrule
Random init & Tokens & 34.33 & 34.33 \\
Random init & CLS & 34.19 & 34.19 \\
Random init & Tokens+CLS & 34.95 & 34.95 \\
\bottomrule
\end{tabular}
\end{table*}

\begin{table*}[h]
\centering
\footnotesize
\caption{Genus-level 1-NN accuracy (\%) on UNITE+INSD (deduplicated Yeast and
Filamentous query pools) for every configuration and representation including the
randomly-initialized encoder baseline. \textbf{Bold} marks the best value in each
column; \underline{underline} marks the second-best.}
\label{tab:appendix_fig2_its}
\setlength{\tabcolsep}{4pt}
\begin{tabular}{llcccc}
\toprule
 & & \multicolumn{2}{c}{Yeast} & \multicolumn{2}{c}{Filamentous} \\
\cmidrule(lr){3-4} \cmidrule(lr){5-6}
Configuration & Representation & Encoder-decoder & Encoder-only & Encoder-decoder & Encoder-only \\
\midrule
(no CLS) & Tokens & 54.94 & 49.81 & 49.11 & 48.85 \\
+CLS & Tokens & 63.50 & 61.79 & 56.63 & 55.93 \\
+CLS & CLS & 67.87 & 66.73 & 59.95 & 56.28 \\
+CLS & Tokens+CLS & 64.26 & 62.17 & 56.86 & 56.09 \\
+CLS+Binary & Tokens & 66.54 & 65.02 & 57.27 & 56.76 \\
+CLS+Binary & CLS & \textbf{73.19} & \underline{67.68} & \textbf{63.07} & \underline{60.27} \\
+CLS+Binary & Tokens+CLS & 66.35 & 65.21 & 57.27 & 57.02 \\
+CLS+Triplet(hard) & Tokens & 66.35 & 64.26 & 57.11 & 56.98 \\
+CLS+Triplet(hard) & CLS & \underline{71.86} & \textbf{69.58} & \underline{62.12} & \textbf{60.52} \\
+CLS+Triplet(hard) & Tokens+CLS & 66.54 & 64.26 & 57.27 & 57.11 \\
+CLS+Triplet(random) & Tokens & 64.45 & 63.12 & 57.30 & 56.41 \\
+CLS+Triplet(random) & CLS & 66.35 & 65.59 & 55.13 & 58.04 \\
+CLS+Triplet(random) & Tokens+CLS & 64.45 & 63.31 & 57.37 & 56.35 \\
+CLS+CE & Tokens & 65.02 & 65.40 & 57.08 & 57.33 \\
+CLS+CE & CLS & 70.72 & \underline{67.68} & 59.50 & 56.15 \\
+CLS+CE & Tokens+CLS & 65.21 & 65.78 & 57.37 & 57.40 \\
\midrule
Random init & Tokens & 42.97 & 42.97 & 47.26 & 47.26 \\
Random init & CLS & 40.87 & 40.87 & 46.52 & 46.52 \\
Random init & Tokens+CLS & 41.63 & 41.63 & 46.56 & 46.56 \\
\bottomrule
\end{tabular}
\end{table*}

\FloatBarrier
\section{H. Jumbo CLS Token Results}
\label{app:jumbo}

We evaluate Jumbo CLS tokens~\citep{fuller2025jumbo} with $J\in\{3,6\}$ and an MLP
expansion factor of $1\times$. On BIOSCAN-5M, the best Jumbo result is obtained with
$J=6$ using the CLS representation (69.38\% 1-NN accuracy and 74.53\% AMI), below the
standard CLS baseline (80.65\% and 75.79\%). On UNITE+INSD, the best Jumbo model also
uses $J=6$, reaching 65.02\% on Yeast and 56.25\% on Filamentous, compared with 73.19\%
and 63.07\% for standard CLS. Thus, Jumbo CLS does not improve performance in the
evaluated configurations. \autoref{tab:appendix_jumbo} and \autoref{tab:appendix_jumbo_its} report
the full results.

\begin{table}[h]
\centering
\small
\caption{BIOSCAN-5M Jumbo CLS token supplemental results (MLP expansion $1\times$).
Genus-level 1-NN accuracy (\%) and zero-shot BIN reconstruction AMI (\%),
for the encoder-decoder, +CLS+CE configuration. \textbf{Bold} marks the best (highest) value in
each column; \underline{underline} marks the second-best.}
\label{tab:appendix_jumbo}
\setlength{\tabcolsep}{5pt}
\begin{tabular}{llcc}
\toprule
$J$ & Repr & 1-NN Acc.\ (\%) & ZSC AMI (\%) \\
\midrule
3 & Tokens  & \underline{62.54} & 73.50 \\
3 & CLS     & 40.43 & 72.19 \\
3 & Tokens+CLS & 62.40 & \underline{73.52} \\
6 & Tokens  & 58.86 & 72.63 \\
6 & CLS     & \textbf{69.38} & \textbf{74.53} \\
6 & Tokens+CLS & 58.57 & 72.72 \\
\bottomrule
\end{tabular}
\end{table}

\begin{table}[h]
\centering
\small
\caption{UNITE+INSD Jumbo CLS token supplemental results (MLP expansion $1\times$).
Genus-level 1-NN accuracy (\%) on the deduplicated Yeast and Filamentous
query pools, for the encoder-decoder, +CLS+Binary configuration. \textbf{Bold} marks the best
(highest) value in each column; \underline{underline} marks the second-best.}
\label{tab:appendix_jumbo_its}
\setlength{\tabcolsep}{5pt}
\begin{tabular}{llcc}
\toprule
$J$ & Repr & Yeast 1-NN (\%) & Filamentous 1-NN (\%) \\
\midrule
3 & Tokens  & 47.53 & 43.91 \\
3 & CLS     & \underline{63.12} & \underline{54.43} \\
3 & Tokens+CLS & 48.29 & 44.58 \\
6 & Tokens  & 39.16 & 37.69 \\
6 & CLS     & \textbf{65.02} & \textbf{56.25} \\
6 & Tokens+CLS & 43.35 & 41.45 \\
\bottomrule
\end{tabular}
\end{table}

\FloatBarrier
\section{I. Softmax vs.\ Uniform KNN Across All Configurations}
\label{app:softmax_all}

\autoref{tab:appendix_softmax_bioscan} and \autoref{tab:appendix_softmax_its} extend the main-text stability analysis to all ten configurations for BIOSCAN-5M and UNITE+INSD, respectively. We use Tokens for models without CLS and CLS otherwise, reporting peak accuracy, mean, standard deviation, and range across $k \in \{1,3,5,7,10,15,20,25,50\}$. The best configurations remain encoder-decoder +CLS+CE for BIOSCAN-5M and encoder-decoder +CLS+Binary for UNITE+INSD.
\begin{table*}[h]
\centering
\footnotesize
\caption{BIOSCAN-5M stability statistics for all ten model configurations, under uniform
and softmax voting, across $k \in \{1,3,5,7,10,15,20,25,50\}$. Mean is the average
accuracy across all nine values of $k$ tested. \textbf{Bold} marks the best Peak/Mean
accuracy in each column; \underline{underline} marks the second-best.}
\label{tab:appendix_softmax_bioscan}
\setlength{\tabcolsep}{4pt}
\begin{tabular}{llcccccccc}
\toprule
& & \multicolumn{4}{c}{Uniform} & \multicolumn{4}{c}{Softmax} \\
\cmidrule(lr){3-6} \cmidrule(lr){7-10}
Arch & Configuration & Peak (\%) & Mean (\%) & Std.\ (pp) & Range (pp) & Peak (\%) & Mean (\%) & Std.\ (pp) & Range (pp) \\
\midrule
Encoder-decoder & (no CLS)      & 66.73 ($k{=}1$) & 60.51 & 4.87 & 17.14 & 66.73 ($k{=}1$) & 62.96 & 1.97 & \phantom{0}6.75 \\
Encoder-decoder & +CLS          & 47.53 ($k{=}1$) & 41.72 & 4.22 & 14.26 & 47.53 ($k{=}1$) & 43.79 & 2.65 & \phantom{0}8.25 \\
Encoder-decoder & +CLS+Binary   & 74.76 ($k{=}\phantom{0}1$) & \underline{66.61} & 6.29 & 21.38 & 74.76 ($k{=}\phantom{0}1$) & \underline{74.02} & 0.74 & \phantom{0}2.33 \\
Encoder-decoder & +CLS+Triplet  & \underline{74.85} ($k{=}\phantom{0}1$) & 63.47 & 7.65 & 24.41 & \underline{74.85} ($k{=}\phantom{0}1$) & 70.00 & 3.49 & 10.34 \\
Encoder-decoder & +CLS+CE       & \textbf{80.65} ($k{=}\phantom{0}1$) & \textbf{72.55} & 7.00 & 23.35 & \textbf{80.65} ($k{=}\phantom{0}1$) & \textbf{80.53} & 0.09 & \phantom{0}0.27 \\
\midrule
Encoder-only & (no CLS)      & 51.27 ($k{=}1$) & 43.66 & 5.89 & 18.29 & 51.27 ($k{=}1$) & 46.85 & 2.67 & \phantom{0}8.66 \\
Encoder-only & +CLS          & 47.53 ($k{=}1$) & 39.59 & 5.79 & 19.00 & 47.53 ($k{=}1$) & 41.20 & 4.48 & 15.08 \\
Encoder-only & +CLS+Binary   & 67.26 ($k{=}1$) & 60.44 & 5.41 & 18.76 & 67.26 ($k{=}1$) & 63.41 & 3.01 & 11.08 \\
Encoder-only & +CLS+Triplet  & 66.02 ($k{=}1$) & 57.95 & 5.56 & 19.02 & 66.02 ($k{=}1$) & 58.47 & 4.80 & 17.23 \\
Encoder-only & +CLS+CE       & 71.67 ($k{=}1$) & 63.30 & 6.80 & 21.58 & 71.67 ($k{=}1$) & 68.71 & 2.61 & \phantom{0}8.36 \\
\bottomrule
\end{tabular}
\end{table*}

\begin{table*}[h]
\centering
\footnotesize
\caption{UNITE+INSD stability statistics for all ten model configurations, on the
deduplicated Yeast and Filamentous query pools, under uniform and softmax voting.
Mean is the average accuracy across the $k$ values tested. \textbf{Bold} marks the best
Peak/Mean accuracy in each column; \underline{underline} marks the second-best.}
\label{tab:appendix_softmax_its}
\setlength{\tabcolsep}{2.5pt}
\begin{tabular}{lllcccccccc}
\toprule
& & & \multicolumn{4}{c}{Uniform} & \multicolumn{4}{c}{Softmax} \\
\cmidrule(lr){4-7} \cmidrule(lr){8-11}
Test set & Arch & Configuration & Peak (\%) & Mean (\%) & Std.\ (pp) & Range (pp) & Peak (\%) & Mean (\%) & Std.\ (pp) & Range (pp) \\
\midrule
\multirow{10}{*}{Yeast}
  & Encoder-decoder & (no CLS)      & 56.84 ($k{=}3$) & 55.13 & 1.22 & \phantom{0}3.42  & 57.79 ($k{=}\phantom{0}5$)  & 54.24 & 2.83 & 10.07 \\
  & Encoder-decoder & +CLS                    & 69.39 ($k{=}3$) & 64.98 & 4.62 & 14.07 & 71.10 ($k{=}\phantom{0}5$)  & 68.27 & 2.48 & \phantom{0}8.55 \\
  & Encoder-decoder & +CLS+Binary             & \textbf{74.90} ($k{=}7$) & \underline{72.05} & 3.10 & 10.45 & \textbf{76.43} ($k{=}25$) & \textbf{75.14} & 1.12 & \phantom{0}3.24 \\
  & Encoder-decoder & +CLS+Triplet            & 73.00 ($k{=}3$) & 70.78 & 2.22 & \phantom{0}7.60  & 73.57 ($k{=}\phantom{0}3$)  & 72.09 & 1.07 & \phantom{0}3.99 \\
  & Encoder-decoder & +CLS+CE       & \underline{73.57} ($k{=}7$) & \textbf{72.57} & 1.13 & \phantom{0}2.85  & \underline{75.29} ($k{=}10$) & \underline{73.68} & 1.35 & \phantom{0}4.57 \\
  & Encoder-only & (no CLS)      & 51.14 ($k{=}7$) & 49.76 & 0.86 & \phantom{0}2.28  & 51.71 ($k{=}10$) & 50.74 & 0.65 & \phantom{0}1.90 \\
  & Encoder-only & +CLS          & 67.11 ($k{=}5$) & 66.06 & 1.28 & \phantom{0}3.23  & 67.68 ($k{=}\phantom{0}5$)  & 65.38 & 1.45 & \phantom{0}4.37 \\
  & Encoder-only & +CLS+Binary   & 69.58 ($k{=}3$) & 68.44 & 0.75 & \phantom{0}1.90  & 69.96 ($k{=}\phantom{0}7$)  & 68.57 & 1.30 & \phantom{0}3.80 \\
  & Encoder-only & +CLS+Triplet  & 71.10 ($k{=}3$) & 70.01 & 0.64 & \phantom{0}1.52  & 70.72 ($k{=}\phantom{0}3$)  & 69.39 & 1.14 & \phantom{0}3.23 \\
  & Encoder-only & +CLS+CE       & 67.68 ($k{=}1$) & 66.68 & 0.82 & \phantom{0}1.90  & 69.20 ($k{=}\phantom{0}3$)  & 67.43 & 0.83 & \phantom{0}3.23 \\
\midrule
\multirow{10}{*}{Filamentous}
  & Encoder-decoder & (no CLS)      & 49.11 ($k{=}1$) & 47.70 & 0.97 & \phantom{0}2.59  & 49.11 ($k{=}\phantom{0}3$)  & 47.05 & 1.41 & \phantom{0}4.18 \\
  & Encoder-decoder & +CLS                    & 61.19 ($k{=}5$) & 59.76 & 1.16 & \phantom{0}3.76  & 61.64 ($k{=}\phantom{0}5$)  & 60.96 & 0.54 & \phantom{0}1.69 \\
  & Encoder-decoder & +CLS+Binary             & \underline{63.36} ($k{=}7$) & \textbf{62.50} & 0.98 & \phantom{0}3.32  & \textbf{65.21} ($k{=}50$) & \textbf{64.01} & 0.71 & \phantom{0}2.17 \\
  & Encoder-decoder & +CLS+Triplet            & \textbf{63.58} ($k{=}5$) & \underline{61.94} & 1.34 & \phantom{0}4.97  & \underline{63.71} ($k{=}\phantom{0}5$)  & \underline{62.92} & 0.77 & \phantom{0}2.61 \\
  & Encoder-decoder & +CLS+CE       & 60.78 ($k{=}7$) & 60.21 & 0.46 & \phantom{0}1.28  & 61.83 ($k{=}50$) & 61.12 & 0.67 & \phantom{0}2.26 \\
  & Encoder-only & (no CLS)      & 49.33 ($k{=}3$) & 48.73 & 0.41 & \phantom{0}1.12  & 50.54 ($k{=}\phantom{0}3$)  & 48.95 & 1.04 & \phantom{0}4.05 \\
  & Encoder-only & +CLS          & 57.43 ($k{=}3$) & 56.79 & 0.43 & \phantom{0}1.15  & 57.59 ($k{=}10$) & 56.86 & 0.45 & \phantom{0}1.40 \\
  & Encoder-only & +CLS+Binary   & 60.36 ($k{=}5$) & 60.16 & 0.17 & \phantom{0}0.44  & 61.07 ($k{=}\phantom{0}7$)  & 60.24 & 0.53 & \phantom{0}1.76 \\
  & Encoder-only & +CLS+Triplet  & 61.48 ($k{=}5$) & 60.88 & 0.44 & \phantom{0}1.08  & 62.15 ($k{=}\phantom{0}7$)  & 60.96 & 0.81 & \phantom{0}2.90 \\
  & Encoder-only & +CLS+CE       & 56.89 ($k{=}5$) & 56.66 & 0.30 & \phantom{0}0.74  & 58.67 ($k{=}50$) & 57.69 & 0.65 & \phantom{0}2.52 \\
\bottomrule
\end{tabular}
\end{table*}
\FloatBarrier

\section{J. Baseline Models}
\label{app:baseline_models}

For evaluation, we used the pretrained checkpoints and public codebases of each
baseline directly. DNA foundation model
baselines were loaded from Hugging Face's Model Hub:
\begin{itemize}
    \item DNABERT-2: \url{https://huggingface.co/zhihan1996/DNABERT-2-117M}
    \item DNABERT-S: \url{https://huggingface.co/zhihan1996/DNABERT-S}
    \item Nucleotide Transformer: \url{https://huggingface.co/InstaDeepAI/nucleotide-transformer-v2-50m-multi-species}
    \item HyenaDNA-tiny: \url{https://huggingface.co/LongSafari/hyenadna-tiny-1k-seqlen}
    \item Caduceus-PS-1k: \url{https://huggingface.co/kuleshov-group/caduceus-ps_seqlen-1k_d_model-256_n_layer-4_lr-8e-3}
    \item GENA-LM (ModernGENA): \url{https://huggingface.co/AIRI-Institute/moderngena-base}
    \item BarcodeBERT: \url{https://huggingface.co/bioscan-ml/BarcodeBERT}
\end{itemize}

Two baselines are not distributed as Hugging Face checkpoints and were instead
obtained from their public GitHub repositories:
\begin{itemize}
    \item BarcodeMamba+~\cite{gao2024barcodemamba,gao2025barcodemambaplus} (both the regular and large variants):
    \url{https://github.com/bioscan-ml/BarcodeMamba-dev}
    \item MycoAI-BERT and MycoAI-CNN~\cite{MycoAI}:
    \url{https://github.com/MycoAI/MycoAI}
\end{itemize}

\end{document}